\documentclass[11pt,a4paper]{article}
\pdfoutput=1
\usepackage{jcappub}
\usepackage{natbib}
\usepackage{graphicx}
\usepackage[english]{babel}
\usepackage{color}
\usepackage{multirow}

\graphicspath{{./figs/}{./skymaps/}}

\title{Galactic synchrotron emission from WIMPs at radio frequencies}

\author[a]{Nicolao Fornengo}
\author[b]{Roberto A. Lineros}
\author[a]{Marco Regis}
\author[b]{Marco Taoso}

\affiliation[a]{Dipartimento di Fisica Teorica, Universit\`{a} di Torino, Istituto Nazionale di Fisica Nucleare, via P. Giuria 1, I--10125 Torino, Italy}
\affiliation[b]{IFIC, CSIC--Universidad de Valencia, Ed.~Institutos, Apdo.~Correos 22085, E--46071 Valencia, Spain, and MultiDark fellow}

\emailAdd{fornengo@to.infn.it}
\emailAdd{rlineros@ific.uv.es}
\emailAdd{regis@to.infn.it}
\emailAdd{taoso@ific.uv.es }


\arxivnumber{1110.4337}

\abstract{
Dark matter annihilations in the Galactic halo inject relativistic electrons and positrons which in turn generate a synchrotron radiation when interacting with the galactic magnetic field.
We calculate the synchrotron flux for various dark matter annihilation channels, masses, and astrophysical assumptions in the low--frequency range and compare our results with radio surveys from $22$ MHz to $1420$ MHz. 
We find that current observations are able to constrain particle dark matter with ``thermal'' annihilation cross-sections, i.e. $( \sigma v ) =  3 \times 10^{-26}$~cm$^3$~s$^{-1},$ and masses $M_{\rm DM} \lesssim 10 \, {\rm GeV}$.
We discuss the dependence of these bounds on the astrophysical assumptions, namely galactic dark matter distribution, cosmic rays propagation parameters, and structure of the galactic magnetic field. 
Prospects for detection in future radio surveys are outlined.

}

\date{\today}

\begin{document}
\maketitle

\section{Introduction}
\label{sec:introduction}

The presence in the Universe of a sizable amount of Dark Matter (DM) has puzzled our comprehension about both fundamental physics and the cosmos.
%
One of the most natural solutions to this problem is that DM is formed by relic particles in the form of Weakly Interactive Massive Particles (WIMP).
WIMPs can in fact easily thermalize in the early Universe and non--relativistically decouple from the plasma with the right abundance when they have a thermally--averaged annihilation cross--sections of the order of $( \sigma v ) =  3 \times 10^{-26}$~cm$^3$~s$^{-1}$. Dynamically they are then cold enough to allow structure formation as required by the cold dark matter paradigm.
%

Several ways have been proposed on how to search for WIMP dark matter. 
Direct detection is the most straight technique.
Recent results from DAMA~\cite{DAMA:2008,Bernabei:2010mq}, CoGeNT~\cite{Fox2011CoGeNT} and CRESST-II~\cite{CRESST:2011} all point toward a possible signal induced by a DM with mass in the 10 GeV range. 
A different approach is to search for WIMP annihilations or decays that could take place in many astrophysical environments like the galactic halo, dwarf galaxies, cluster of galaxies, the Sun and the Earth or even during the early dark ages of the Universe.
An interesting result comes from the FERMI--LAT Collaboration, that recently
presented significant bounds on dark matter annihilations into gamma--rays occurring
in satellite galaxies of the Milky Way~\cite{Ackermann:2011}. 
From this analysis, WIMP candidates with ``thermal'' annihilation cross-sections (i.e. $( \sigma v ) =  3 \times 10^{-26}$~cm$^3$~s$^{-1},$) and inducing a flux of $\gamma$--rays from $\pi^0$--decay (i.e., annihilating into quarks or $\tau$--leptons) are strongly constrained for masses below few tens of GeV~\cite{Ackermann:2011,GeringerSameth:2011iw}.
On the other hand, WIMP candidates annihilating into light leptons are much less strongly constrained.
%

In this work, we focus our attention on the Galactic synchrotron emission from DM annihilations and we concentrate on observations in the frequency range from 22 to 1420 MHz. 
We extend previous analyses \cite{Borriello:2009,Borriello:2009a,Delahaye:2011jn,Regis:2009md,Siffert:2010cc} to cover the very low--frequency
range, specifically we will consider radio emission at 22 and 45 MHz. Such
low radio frequencies have not been addressed in the past in connection with DM studies, and we will show that they may be quite adequate in the search for DM,
especially for low--mass WIMPs which are currently under deep scrutiny due to the direct detection results quoted above.
Indeed, for a magnetic field of $\mu$G strength, as typical in the Galaxy, synchrotron emission below GHz--frequency is generated by electrons and positrons with energies well below 10 GeV.
In particular, we will show that WIMP candidates annihilating into light leptons with a ``thermal'' annihilation cross-sections can be strongly constrained for masses $M_{\rm DM} \lesssim 10 \, {\rm GeV}$.

Radio emission arising from DM annihilation have instead been addressed in Refs.~\cite{Borriello:2009,Borriello:2009a,Delahaye:2011jn,Regis:2009md,Siffert:2010cc} by using higher frequency surveys and by studying heavier dark matter candidates (more suitable to explain the PAMELA positron excess~\cite{Adriani:2009a}), with a particular focus on the Galactic Center by Refs.~\cite{Bertone:2002je,Bertone:2008xr,Boehm:2010kg,Crocker:2010gy,Regis:2008ij,Bergstrom:2008ag}, or in order to reproduce the so called ``WMAP Haze'' in Refs.~\cite{Cumberbatch:2009ji,Dobler:2007wv,Linden:2010eu,Hooper:2007kb}. DM searches with radio observations can also be pursued 
with the cosmological diffuse radio background, using both intensity \cite{Fornengo:2011cn,Fornengo:2011cnl} and anisotropies informations \cite{Zhang:2008rs,Fornengo:2011c}.

%
The outline of this work is the following.
%
In section~\ref{sec:darkmatter}, we describe general aspects about galactic DM and we
discuss our approach to the dark matter profiles used in our calculation.
In section~\ref{sec:cosmicrays}, we review electrons and positron propagation in the
galactic environment and discuss the main sources of uncertainty which will then represent
a key element of discussion for the predicted radio fluxes.

In section~\ref{sec:synchrotron}, we discuss the synchrotron emission from a galactic
population of electrons and describe models for the Galactic Magnetic Field (GMF) distribution. 
%
Sections~\ref{comparison} and \ref{constraints} are devoted to the results on the radio emission from DM annihilations and to comparison with experimental data.
We specifically adopt available temperature skymaps at 5 frequencies: 22, 45, 408, 820, 1420~MHz ~\cite{1999A&AS..137....7R,Guzman2010,1982A&AS...47....1H,1972A&AS....5..263B,1986A&AS...63..205R}. 
From this comparison we then derive bounds on the DM particle properties (annihilation cross sections and channels, DM mass). 
We discuss the impact of the uncertainties on the galactic halo modeling, electron propagation, size and shape of the GMF.

In Section~\ref{prospects}, we analyze possible traces of WIMP annihilation at 22 and 45 MHz after subtraction of a template based on the Haslam map~\cite{1982A&AS...47....1H} for the astrophysical component.
In Section~\ref{conclusions} we summarize our conclusions.

\section{Dark Matter distribution}
\label{sec:darkmatter}

N--body simulations are currently the best tool to study the formation and evolution of cosmic structures. 
Galactic DM halos obtained by these simulations contain a smooth halo component together with a population of clumps and other possible unvirialized structures, such as streams.

The radial dark matter distribution of a Milky Way--like DM halo obtained by the Via Lactea II (VLII) simulation~\cite{Diemand2008} is well captured by a Navarro--Frenk--White (NFW) density profile~\cite{NavarroAstrophys.J.490:493-5081997}:
\begin{equation}
	\label{nfw}
	\rho^{\rm{NFW}}(r) = \frac{\rho_s}{\frac{r}{r_s}\left(1+\frac{r}{r_s}\right)^2}
\end{equation}
where $r$ is the distance from the center of the DM halo and the values of the parameters $\rho_s$ and $r_s$ are $r_s = 21 {\rm kpc}$ and $\rho_s=0.31$~GeV~cm$^{-3}$ 
\cite{Pieri:2009je}. The local DM density is then $\rho_l = 0.43$~GeV cm$^{-3}$.
The results of the Aquarius simulation are instead better fitted  by a slightly shallower Einasto profile~\cite{Springel2008,Springel2008a}. 
%
In the determination of the NFW parameters we closely follow the analysis of
Ref. \cite{Pieri:2009je}, where a detailed study of the dark matter density
profiles arising from numerical simulations has been performed, and where the
relevant profile parameters have been determined. With the above values, the
halo mass within the virial radius is correctly reproduced \cite{Pieri:2009je}.

%
It is however unclear whether these profiles properly describe the DM distribution in the innermost regions of the halo, which are not accessible to numerical simulations due to their finite resolutions.
Moreover, the feedback of baryons on the DM distribution, not considered in the simulations
cited above, can be relevant in the inner regions of the Milky--Way.
In particular, the condensation of gas and stars can produce an adiabatic contraction
of the DM halo, increasing the central DM density~\cite{1986ApJ...301...27B,1987ApJ...318...15R}.
This effect has been confirmed by recent hydrodynamical simulations which include the presence of gas and stars
(see~Ref. \cite{Gnedin2011} and refereces therein).
Also the effect of the central super massive black hole on the surrounding DM distribution
is uncertain since it may either induce an adiabatic contraction of the DM halo or
destroy the central DM cusp during its spirraling at the center of the halo.
The central DM cusp can also be erased by hierarchical mergers occurred to the Mily Way halo
during its recent evolution~\cite{Ullio:2001fb,Read:2003,Vasiliev:2007}.
Current microlesing and dynamical observations are consistent with the NFW and Einasto profiles found in numerical simulations and can be used to constraint combinations of local DM densities and inner slopes of the DM profiles, ruling out extremely cuspy DM distributions~\cite{Iocco2011}. 
Nevertheless, cored profiles, despite disfavored by simulations, are observationally viable and in some cases preferred over cuspy ones, like for observations of dwarf spirals and low-surface brightness galaxies~\cite{Gentile2004}.

In our analysis, in order to bracket the uncertainty arising from the
shape of the DM profile, we consider the NFW profile discussed above and an isothermal profile, the latter used a conservative case of a cored DM distribution:
\begin{equation}
	\rho^{\rm{ISO}}(r)=\rho_0\; \frac{r_a^2}{r^2+r_a^2}
	\label{isothermal}
\end{equation}
with a core radius $r_a = 5 {\rm~kpc}$. 
The normalization $\rho_0$ is chosen such that
the same dark matter density of the NFW profile of Eq. (\ref{nfw}) occurs at the Sun position,
which we take at $r_{\odot} = 8 {\rm ~kpc}$. Therefore $\rho_0 = 1.53$~GeV~cm$^{-3}$.

In addition to the smooth density distribution discussed above, the Milky--Way
may host a population of DM clumps which would give rise to an additional contribution to the electron/positron flux induced by DM annihilations.
Unfortunately, the properties of this population of subhalos, which are inferred from simulations, are affected by many uncertainties. 
Since our study is meant to derive
bounds on DM annihilation in the Galaxy, we choose not to include substructures
in our analysis. 
In this case, the constraints we derive in Sec.~\ref{comparison} are
conservative, since the effect of the presence of clumps is to boost up the radio fluxes produced by the smooth halo. Nevertheless, we have made an estimate of a potential
effect of clumps on our results, and we found that their inclusion would not not affect the main results of this paper, {\em i.e.} the bounds on the DM annihilation cross section.
To perform this check we have considered a population of clumps modeled following the results of the VLII simulation~\cite{Kuhlen2007,Anderson2010,Kuhlen2008} and using the prescriptions
of Ref. \cite{Pieri:2009je}. We found that in a wide region ($\approx 30$ degrees) around
the galactic center, the average radio flux from the population of clumps can be safely
neglected as compared to the radio flux produced by the smooth halo. This would be even
more so for the Aquarius simulation \cite{Pieri:2009je} , and
occurs because the spatial distribution of the clumps in the numerical simulations occur to
be  anti--biased: the majority of the clumps
lie at large radii and their number density is depleted at small galactocentric distances. 
Since the bounds on the dark matter annihilation cross section we will derive in Section~\ref{constraints} will mostly arise from the inner parts of the Galaxy, we can conclude that for our purposes the average contribution of the subhalos can be safely neglected in the present study. The relevant and interesting prospects of detecting individual clumps with radio observations is left for a future study.

\section{Cosmic rays electrons}
\label{sec:cosmicrays}
The propagation of cosmic rays (CR) electrons and positrons is modeled following a 
semi--analytical approach~\cite{Maurin:2001}.
Cosmic rays are assumed to be confined by the Galactic magnetic field inside a propagation zone which
is described by a cylinder centered at the galactic center. Its radius along
the galactic plane $R_g = 20~{\rm kpc}$
approximately matches the radius of the galactic disk.
The value of the vertical half--thickness $(L_z),$ which is instead more uncertain, is
estimated from nuclear cosmic rays observations and takes values from 1 to 15~kpc~\cite{Maurin:2001,Maurin:2002}.
Outside the propagation zone, cosmic rays are no longer confined by the magnetic field
so their density is expected to rapidly drop.
The CR propagation is described by the transport equation for the number density of electrons per unit of energy $(\psi)$: 
\begin{equation}
	\label{eq:transport}
	-K_0 E^\delta \nabla^2 \psi - \frac{\partial}{\partial E} \big(b(E) \psi \big) = q({\bf x},E) \, ,
\end{equation}
where $K_0$ and $\delta$ are parameters related to spatial diffusion, $b(E)$ is the energy loss term, and 
$q({\bf x},E)$ is the source term, that in our case will correspond to DM annihilations.
The energy loss term for electrons takes into account 
the synchrotron and inverse Compton (IC) scattering losses from interactions with magnetic and interstellar radiation fields (ISRF):
\begin{equation}
	b(E) = b_{\rm{ISRF}}(E) + b_{\rm{synch}}(E,B)  \, ,
\end{equation}
where the term related to the ISRF, $b_{\rm{ISRF}}$, is modeled as in Ref.~\cite{Delahaye:2010}.
%
We take under consideration a specific modeling of the various ISRFs and we explicitly include the Klein--Nishina corrections to Compton scattering~\cite{Delahaye:2010}. Nevertheless, these corrections do not produce such important changes in the final result specially for energies lower than 10 GeV, where Compton scattering is in the Thomson regime. 
As a reference value, for electron energies below
a few GeV (which are relevant for light dark matter), 
$b_{\rm{ISRF}} \simeq 0.8 \times 10^{-15}\;(E/{\rm GeV})^{2}$ GeV/s.
%
The energy losses from synchrotron emission depend on the galactic magnetic field intensity $B$ as \cite{Ginzburg:1965}:
\begin{equation}
	b_{\rm{synch}}(E,B) = \frac{4 e^4 B^2}{9 m^4 c^7} \, E^2 \, ,
\end{equation}
where $c$ is the speed of light, $e$ and $m$ denote the electron's charge and mass, respectively.
For a magnetic field of $\mathcal{O}(6 \mu {\rm G})$, the ratio of the synchrotron and IC losses
is:
\begin{equation*}
	b_{\rm{synch}}/b_{\rm ISRF} \simeq 1/2.5 \, ,
\end{equation*}
which indicates that $b_{\rm{synch}}$ does not dominate the full energy loss term.
For this reasons, for conservative values of $B (\lesssim 20~\mu\rm{G})$
the magnetic field can be safety be assumed to be constant inside the propagation zone~\cite{Cirelli:2010,Delahaye:2010}.
This assumption is well justified for what concern the CR propagation while the morphology of the associated synchrotron emission depends on the details of the galactic spatial structure.
%

In the galactic disk ($\sim 200$~pc thick) additional energy losses, like bremsstrahlung and gas ionization, might become relevant, in particular at low electrons energies~\cite{Delahaye:2008}. 
While these effects might be important for CR sources localized on the disk, they are expected to play a minor role in the case of more extended sources, like a DM halo, significantly reducing the synchrotron flux only at latitudes $|b| \lesssim 1^{\circ}$.

%
%
%

The parameters of the propagation model are constrained by 
CR data, notably the Boron over Carbon ratio (B/C) and radioactive isotopes~\cite{Maurin:2001,Maurin:2002,Putze:2010}.
Let us mention that the resulting viable parameter space is also consistent with the secondary positrons data~\cite{Delahaye:2008}.
In this study, we use the following three propagation benchmarks: MIN, MED, and MAX~(Table~\ref{tab:models_res}). 
These models are fully compatible with B/C and antiproton--proton ratio ($\bar{p}/p$) observations, 
describing the extremes behaviors (MIN and MAX) and best fit (MED) of these observables~\cite{Donato:2001,Moskalenko:2002}.

Recent efforts to further constraint the space of the propagation parameters have focused to ``alternative'' observables, like synchrotron total intensity~\cite{Bringmann:2011,Strong:2011} and anisotropies~\cite{Regis:2011}, and gamma rays~\cite{Delahaye:2011}.

\subsection{Galactic electron population from dark matter}

\begin{table}[tb]
	\centering
	\begin{tabular}{|c||c|c|c|}
	\hline
	\multirow{2}{*}{\bf Model}  & \multicolumn{3}{c|}{\bf propagation parameters} \\
	& $L_z$ [kpc] & $K_0\left[\frac{{\rm kpc}^2}{\rm Myr}\right]$ & $\delta$ \\
	\hline
	\hline
 	MIN & $1$ & $0.0016$ & 0.85 \\
 	\hline
 	MED & $4$ & $0.0112$ & 0.70 \\
 	\hline
 	MAX & $15$ & $0.0765$ & 0.46 \\
	\hline
\end{tabular}
\caption{\label{tab:models_res}  Benchmark propagation models compatible with B/C and $\bar{p}/p$ observations~\cite{Maurin:2001,Maurin:2002,Donato:2004}.}
\end{table}

In addition to standard CR astrophysical sources, dark matter annihilations in our Galaxy
could provide an exotic source of CR electrons and positrons.
The resulting electrons number density is computed from the transport equation Eq.(\ref{eq:transport})
with the following source term:

\begin{equation}
	\label{eq:source_term}
	q({\bf x},E) = \frac{1}{2} ( \sigma v ) \left( \frac{\rho({\bf x})}{M_{\textnormal{DM}}} \right)^2 \, \frac{dn}{dE}(E) \,,
\end{equation}
where $( \sigma v )$ is the annihilation cross section, $\rho({\bf x})$ is the dark matter density and
$M_{\textnormal{DM}}$ is the dark matter mass. The function
$dn/dE$ is the energy spectrum of electrons/positrons per single annihilation and depends on the particular DM annihilation
channel \cite{Lineros:2008,Delahaye:2007}.

\begin{figure*}[tb]
	\centering
	\includegraphics[width=0.69\textwidth, bb = 15 0 720 450]{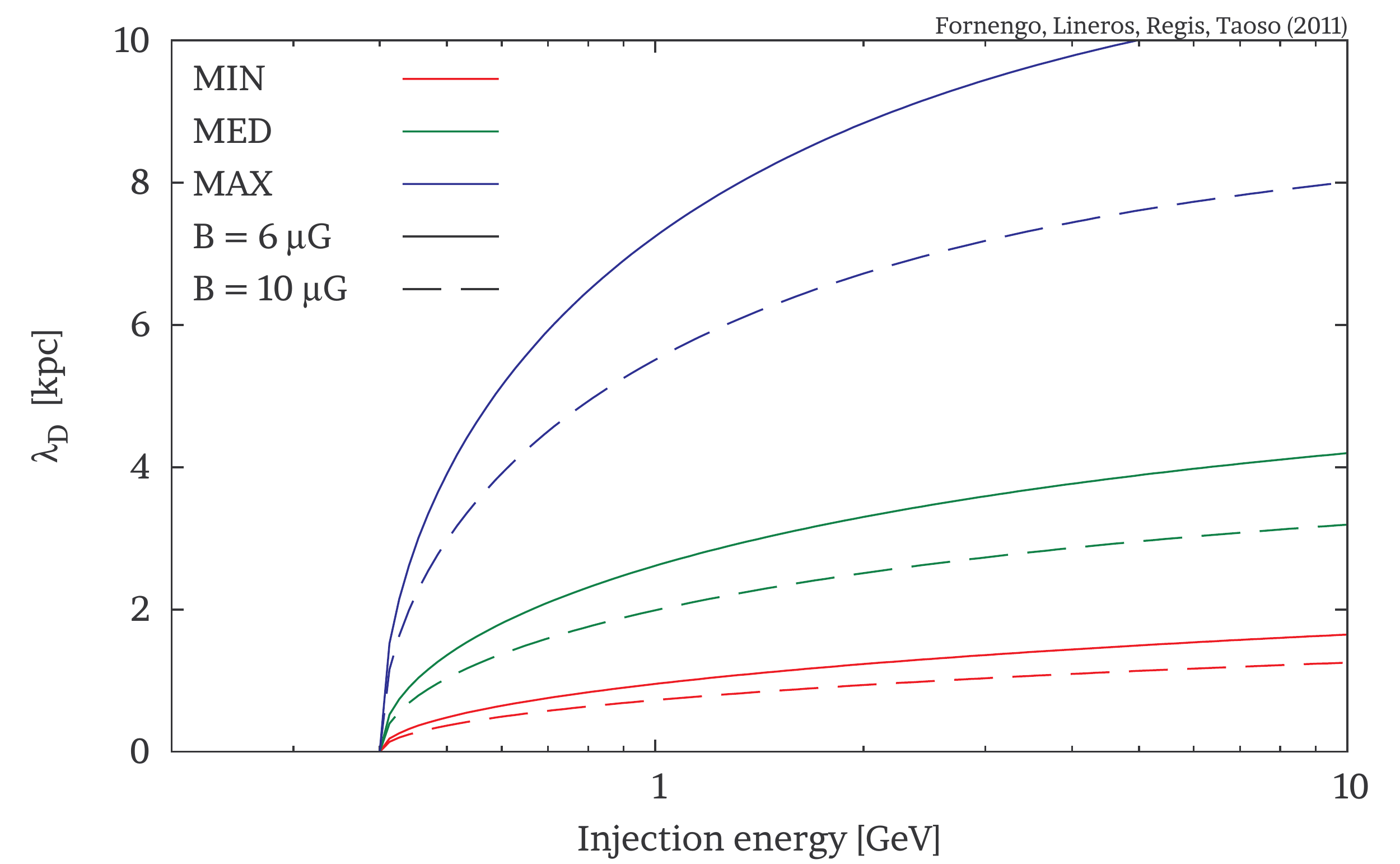}
	\caption{\label{fig:ld} 
Diffusion length $\lambda_D$ for propagated electron of $0.3{\rm~GeV}$ versus injection energy for MIN, MED, MAX propagation models (Table~\ref{tab:models_res}) and magnetic field intensities of $6~\mu{\rm G}$ and $10~\mu{\rm G}$. 
Depending of the propagation model, electrons can easily cover regions larger than ${\rm kpc}^3$.
}
\end{figure*}

The transport equation can be solved using the
formalism of Refs.~\cite{Lineros:2008,Delahaye:2007}, which is based on a decomposition of the source
term in Fourier and Bessel modes, exploiting the axial symmetry of the propagation zone
(for more details see Ref.~\cite{Lineros:2008}).
This method, initially proposed to calculate the electrons/positrons fluxes at the Earth
position,
can be extended to compute the electron density anywhere inside the propagation zone.
The spatial information of the source term is encoded in the so--called \emph{halo function}, $\tilde{I}_{{\bf x}}$, which for a particular position ${\bf x}$ depends only
on one variable $\lambda_{\rm{D}}$:
\begin{equation}
	\tilde{I}_{{\bf x}}(E,E_s) = \tilde{I}_{{\bf x}}(\lambda_{\rm{D}}(E,E_s)) \,,
\end{equation}
where $\lambda_{\rm{D}}$ is the \emph{diffusion length}
\begin{equation}
	\lambda^2_{\rm{D}}(E,E_s) = 4 \int_{E}^{E_s} d\epsilon \, \frac{K_0 \epsilon^{\delta}}{b(\epsilon)} \,,
\end{equation}
%
%
The diffusion length depends on the propagation model and energy losses and it can be understood as the scale length that can be reached by electrons with energy $E$ that were produced with energy $E_s$. 
For propagation models compatible with B/C observations, electrons can cover distances larger than few kpc's becoming mildly insensitive to small--scale variation of energy losses (Fig.\ref{fig:ld}) which can be therefore approximated to be spatially constant.
%
%
%
%
%
%
The source term in Eq.(\ref{eq:source_term}) can be factorized as:
\begin{equation}
	q({\bf x},E) =   f({\bf x}) \times g(E) \,.
\end{equation}
The electron number density is obtained as:
\begin{equation}
	\psi_{e}({\bf x},E) = \frac{1}{b(E)} \, \int_{E}^{M_{\rm{DM}}} dE_s \, \tilde{I}_{\bf x}(E,E_s) \, g(E_s) \, ,
\end{equation}
where the halo function at position ${\bf x}$(= ($r,z$)) is give by:
\begin{equation}
	\label{eq:halosum}
	\tilde{I}_{\bf x}(\lambda_D) = \sum_{i=1}^{\infty} \sum_{n=0}^{\infty} C_{in} \, J_0{\left(\frac{\alpha_{0,i}}{R_g} r\right)} \, \phi_n(z) \, 
\exp\left(\frac{-K_{in}^2 \lambda_D^2}{4} \right) \, .
\end{equation}
The coefficients $C_{in}$ are obtained through the decomposition of $f({\bf x})$ in Fourier and Bessel modes. 
$J_0(x)$ is the first kind bessel function, $\alpha_{0,i}$ is the $i$--th zero of $J_0$ and
$\phi_n$ are the Fourier eigenfunctions:
\begin{equation}
	\phi_n(z) = \left\{ \begin{array}{cc} 
	\sin{\left(k_n z \right)} & n \, \rm{~even}\\
	\cos{\left(k_n z \right)} & n \, \rm{~odd}
	\end{array}\right.
\end{equation}
with $k_n = n\pi/2L_z$.
Finally, $K_{in}$ is given by:
\begin{equation}
	K_{in}^2 = \left(\frac{\alpha_{0,i}}{R_g}\right)^2 + k_n^2 \, .
\end{equation}
For well behaved functions $f({\bf x})$, the halo function is well defined for $\lambda_D = 0$, where it converges to the value:
\begin{equation}
	\tilde{I}_{\bf x}(\lambda_D \rightarrow 0) \rightarrow f({\bf x}) \,
\end{equation}
To ensure the numerical convergence everywhere in the propagation zone and for diffusion lengths larger than 10~pc, we needed to calculate up to $10^7$ coefficients $C_{in}$.\\

\section{Synchrotron radiation}
\label{sec:synchrotron}

\begin{table}[t]
\centering
\begin{tabular}{|c||c|c|c|}
\hline
\multirow{2}{*}{\bf GMF Model}  & \multicolumn{2}{c|}{\bf parameters} \\ 
& $L_m\,[{\rm kpc}]$ & $R_m\,[{\rm kpc}]$ \\
\hline
\hline
I & $\delta L_z$ & $\delta R_g$ \\
\hline
II & $L_z$ & $R_g$  \\
\hline
III & 1 & $R_g$ \\
\hline
IV & \multicolumn{2}{c|}{constant} \\
\hline
\end{tabular}
\caption{
\label{tab:mag_models} 
Parameters of the galactic magnetic field models considered in this study.
We assume that the magnetic field at Sun position ($r_{\odot} = 8 \rm{kpc}$) has a value of ${6~\mu{\rm G}}$.}
\end{table}

Electrons propagating inside the Galaxy interact with the galactic magnetic field (GMF) 
producing synchrotron radiation~\cite{Ginzburg:1965}.
For magnetic field intensity of $\mathcal{O}(\mu{\rm G})$~\cite{Han:2009}, like
in the case of our galaxy and for electrons/positrons of GeV--TeV energies,
the synchrotron emission falls in the MHZ--GHz range~\cite{Goldstein:1970}, i.e. in the radio band.

The synchrotron flux observed at the Earth results from the integration along the line of sight of the synchrotron emissivity $j_{\nu}$:
\begin{equation}
j_{\nu}({\bf x}) = \int dE \, n_e(E,{\bf x}) \frac{dw}{d\nu}({\bf x})
\end{equation}
where
$n_e(E,{\bf x})$ is the  energy spectrum of the electronic number density and $dw/d\nu$ denotes the emission power, which depends on the magnetic field strength $B$({\bf x})
(see Appendix \ref{app:synch} for details).
In our case, the electron spectrum is obtained by solving the diffusion equation for the population of electrons and positrons produced by DM annihilations, as described in the previous section.

The structure of the GMF is still not well understood.
Mainly, it is composed by a regular and a turbulent part, the last one responsible for the diffusive behaviour of the cosmic rays. 
The main methods used to estimate the GMF are Faraday rotation measurements of pulsars 
(more sensitive to the regular part) and diffuse 
synchrotron emission from the non--thermal electron population (sensitive to both components).

In the literature, various parametrization of the GMF distributions
have been proposed~\cite{Han:2006,Evoli:2007,Jansson:2009,Sun:2009,Pshirkov:2011}.
Here, we consider a magnetic field with a cylindrical symmetry and an exponential dependence on the distance on the galactic center $r$:

\begin{equation}
	B(r,z) = B_{0} \exp\left( - \frac{r-r_{\odot}}{R_{m}}\right) f(z) \, .
\end{equation}
The regular component of the GMF typically has $B_0 \sim 2\mu\rm{G}$ and a scale radius $R_{m} \sim 8.5~\rm{kpc}$. 
Its vertical dependence, encoded in $f(z)$, is more uncertain, due to the low number of pulsars at high latitudes.
Uncertainties on the turbulent component of the GMF arise from the fact that the synchrotron emission,
used to estimate it, depends both on the magnetic field intensity and the electrons density.
This degeneracy can be partially broken using local cosmic rays observations. 
%
%

%
An interesting link between the propagation model and the magnetic field is given by the expression of the diffusion term 
of the transport equation Eq. (\ref{eq:transport}) for isotropic turbulences in the
quasi--linear approximations \cite{Berezinskii:1990,Schlickeiser:2002}:
\begin{equation}
	\label{eq:difexpo}
	D(E) = K_0 E^{\delta} \propto r_g^{\delta} \frac{B_{\rm{mean}}^2}{B_{\rm{turb}}^2} \, ,
\end{equation}
%
where $r_g = \mathcal{R}/B_{\rm{mean}}$ is the Larmor radius, which depends on the rigidity ($\mathcal{R}$), $B_{\rm{mean}}$ and $B_{\rm{turb}}$ are the mean and the 
turbulent components of the GMF.
This allows to connect the spatial extension of the GMF with the diffusion coefficient.

In the following we model both the regular and turbulent components of the magnetic field with a double exponential:
\begin{equation}
	B(r,z) = B_0 \exp\left(- \frac{r-r_{\odot}}{R_m} - \frac{|z|}{L_m}\right) \, .
\end{equation}
%
If we assume for the scale parameters $R_m$ and $L_m$ the following relations:
\begin{equation}
L_{m} = \delta L_z \ , \  R_{m} = \delta R_g \,
\end{equation} 
where $\delta$ is one of the diffusion parameters (Eq.~\ref{eq:transport}),
then the resulting diffusion term (Eq.~\ref{eq:difexpo}) will grow exponentially with scales $R_g$ and $L_z$.
%
%
This exponential dependence of the diffusion term on the cylindric coordinates is commonly adopted
in numerical codes like Galprop~\cite{Strong:1998,Moskalenko:1998,Trotta:2010} and Dragon~\cite{Evoli:2008,Bernardo:2009}.
Inside the propagation zone, this parametrization is almost equivalent to a constant
diffusion term.
In addition to this parametrization we have considered a double exponential model with the same
scales as those of the propagation zone and the simplistic case of a constant magnetic field.
As a final additional choice, 
we have  fixed the vertical scale of the magnetic field $L_m$ to 1 kpc, which is much
lower than the height of the diffusion zone in the case of the MED and MAX propagation models. These four parameterizations are summarized in Table \ref{tab:mag_models}.
The strength of the {\it total} magnetic field (turbulent plus regular) at the Sun position is set to ${B=6~\mu{\rm G}},$ which is well consistent with the measurements~\cite{Han:2006,Evoli:2007,Jansson:2009,Sun:2009,Pshirkov:2011}.

\section{Comparison with observations}
\label{comparison}

\begin{table}[tb]
	\centering
	\begin{tabular}{|c||c|c|}
	\hline
	$\nu$ [MHz] & Survey & rms noise [K]\\
	\hline
	\hline
 	22 & DRAO \mbox{\cite{1999A&AS..137....7R}} & 5000\\ %
 	\hline
 	45 & Guzman et al.\mbox{\cite{Guzman2010}}&3500\\
 	\hline
 	408 & Haslam et al.\mbox{\cite{1982A&AS...47....1H}}&0.8\\
	\hline
        820 & Dwingeloo \mbox{\cite{1972A&AS....5..263B}}&1.4\\
	\hline	
	1420 & Stockert \mbox{\cite{1986A&AS...63..205R}}&0.02\\
	\hline
\end{tabular}
\caption{Surveys considered in the analysis. The last column is the rms temperature noise. \label{tab:surveys}  }
\end{table}

\begin{figure*}[t]
	\centering
	\resizebox{\textwidth}{!}{
	\includegraphics[width=0.49\textwidth, bb = 20 20 1130 775]{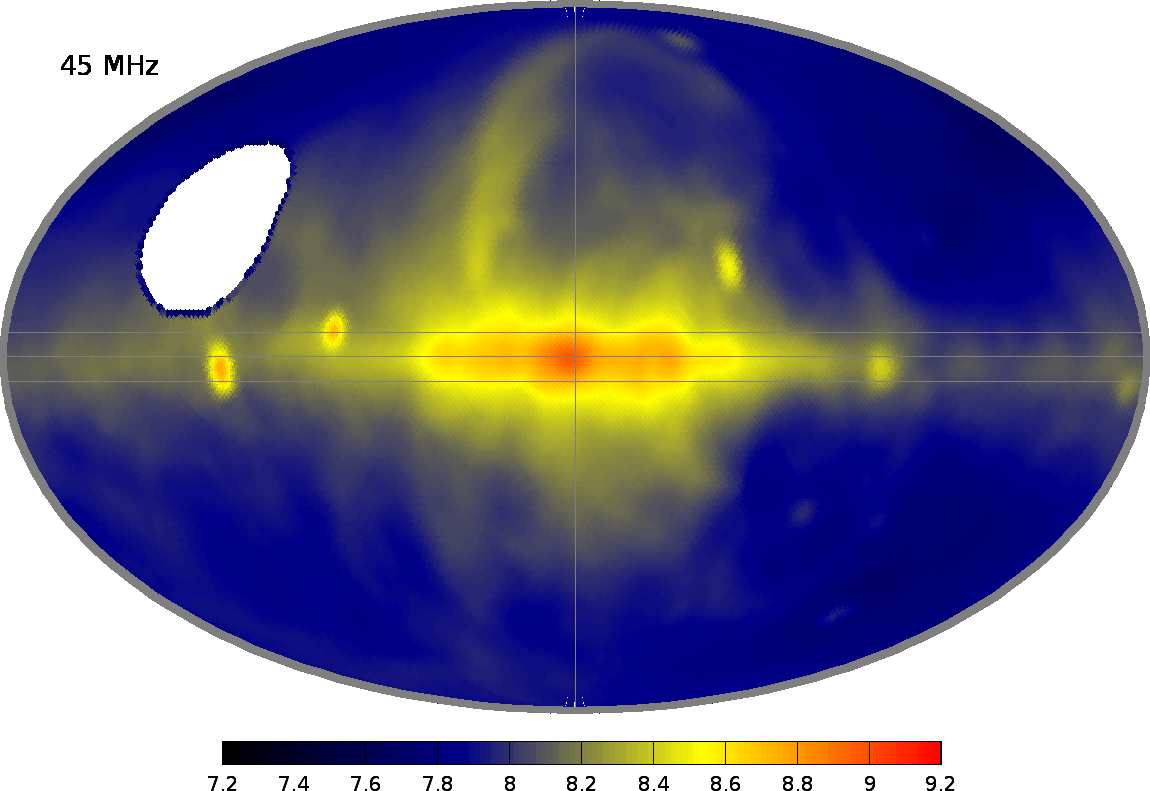}\hspace{2ex}
	\includegraphics[width=0.49\textwidth, bb = 20 20 1130 775]{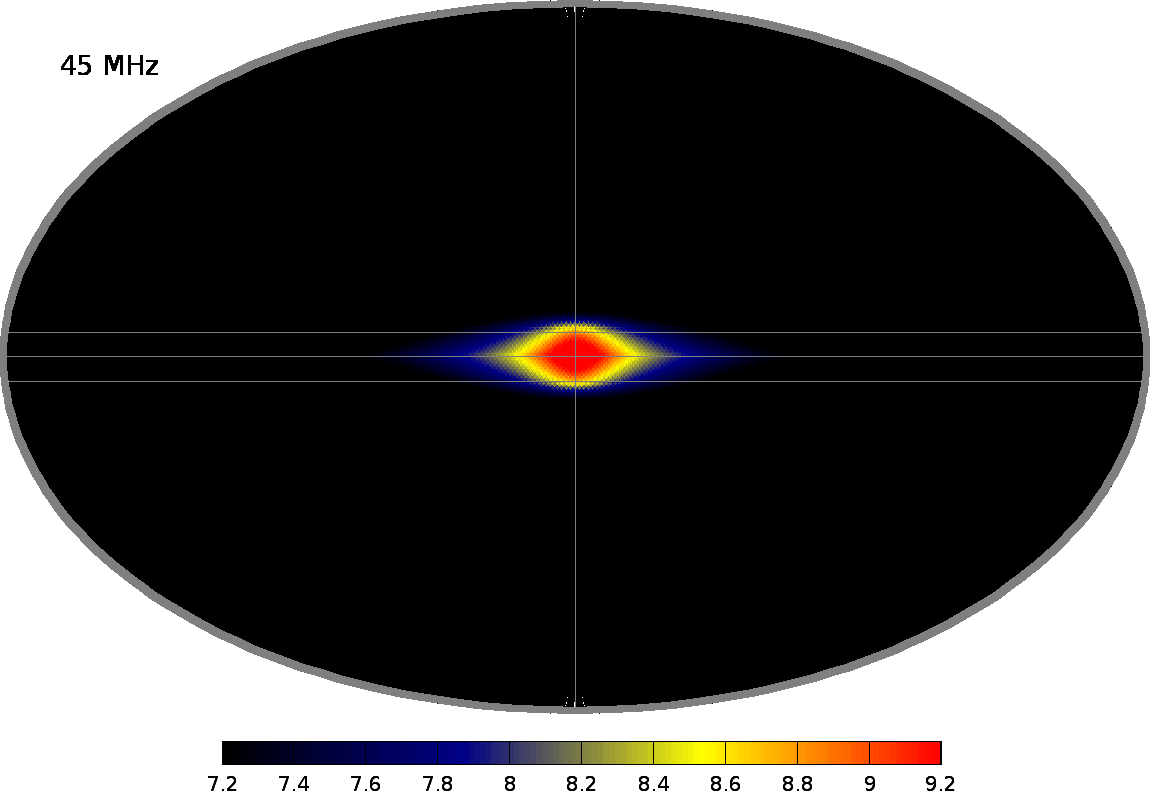}}\\[3ex]
	\resizebox{\textwidth}{!}{
	\includegraphics[width=0.49\textwidth, bb = 20 20 1130 775]{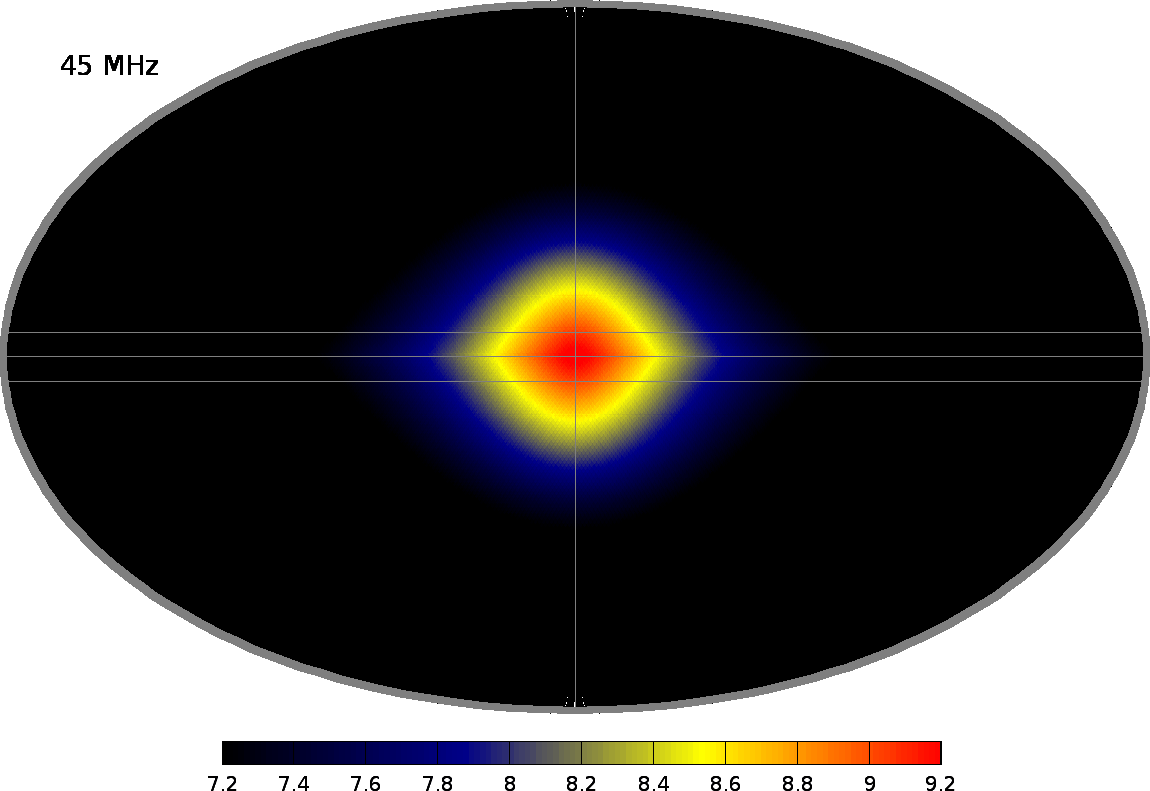}\hspace{2ex}
	\includegraphics[width=0.49\textwidth, bb = 20 20 1130 775]{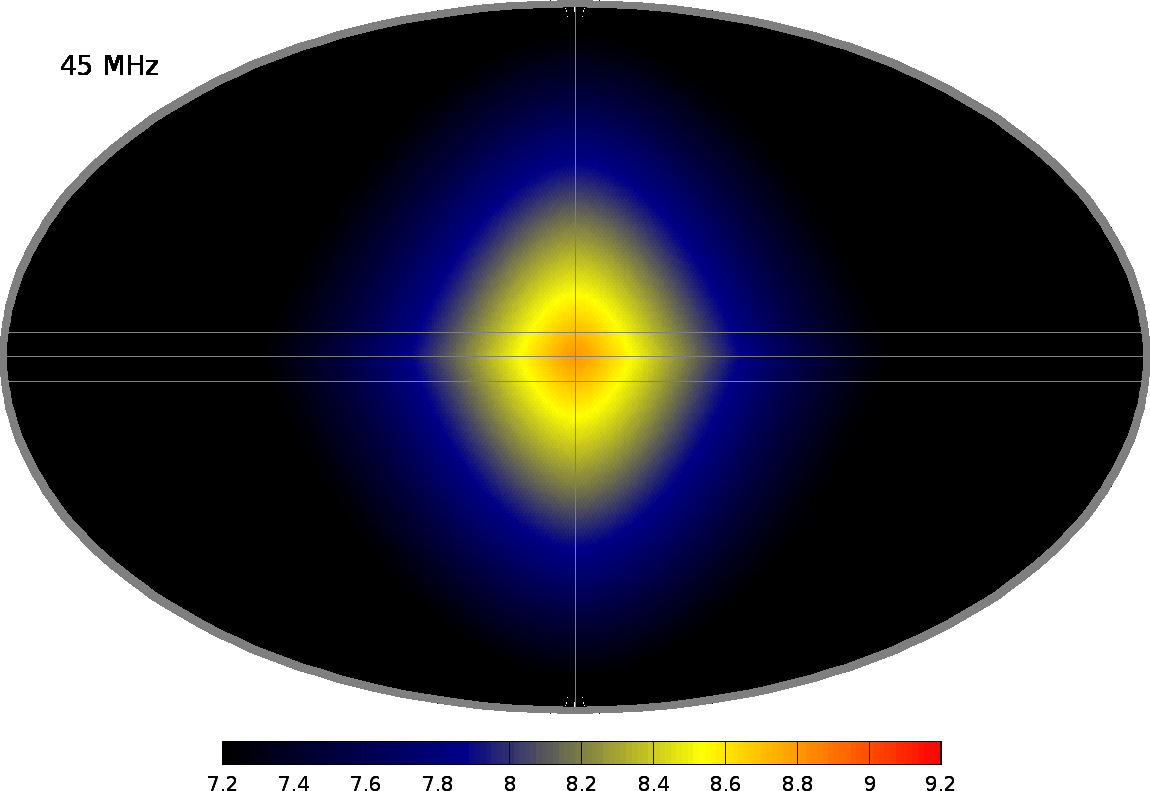}}

	\caption{Temperature skymap of the galactic radio emission observed at 45 MHz (upper left) and synchrotron skymaps at the same frequency from DM annihilations for the three propagation models in table~\ref{tab:models_res}: MIN (upper right), MED (lower left), and MAX (lower right).
The DM synchrotron emission is computed for a 10~GeV DM particle annihilating into muon pairs
with an annihilation cross section $( \sigma v ) = 3 \times 10^{-26}$~cm$^3$~s$^{-1}.$
We adopt the GMF model I and the NFW
profile discussed in section~\ref{sec:darkmatter}. Color scale corresponds to $\log_{10}\Big(T/\rm{K} \ \big(\nu/\rm{MHz}\big) ^{2.5}\Big)$ and it is common for all maps.}
	\label{skymap-mu}
\end{figure*}

Using the formalism explained in the previous sections we have generated skymaps of the synchrotron radiation produced by DM annihilations at different frequencies.
The three examples shown in Fig.~\ref{skymap-mu} are for a $10$~GeV DM particle annihilating into muon pairs and for the three propagation models of Table~\ref{tab:models_res}.
Throughout this Section, for definiteness the DM annihilation cross section
is set to the {\it thermal} value: $( \sigma v ) = 3\times10^{-26}$~cm$^3$~s$^{-1}$. 
In Fig.~\ref{skymap-mu}, the NFW density profile and the GMF model I are used.
%
The morphology of the emission strongly depends on the propagation model, DM spatial profile, and the structure of the GMF.
In particular, the propagation zone height $L_z$ has a large impact, with larger values of $L_z$ leading to more extended emissions.
%

Recently, radio observations have been used in combination with cosmic rays measurements at Earth to constraints the cosmic rays propagation model~\cite{Bringmann:2011,Strong:2011}.
Here, we instead study the bounds on the dark matter annihilations cross section which can be
inferred from current data and we investigate the possibility to detect dark matter annihilations with radio observations.

\begin{figure*}[tb]
	\centering
	\resizebox{0.49\textwidth}{!}{
	\includegraphics[width=0.49\textwidth, bb = 20 20 1130 775]{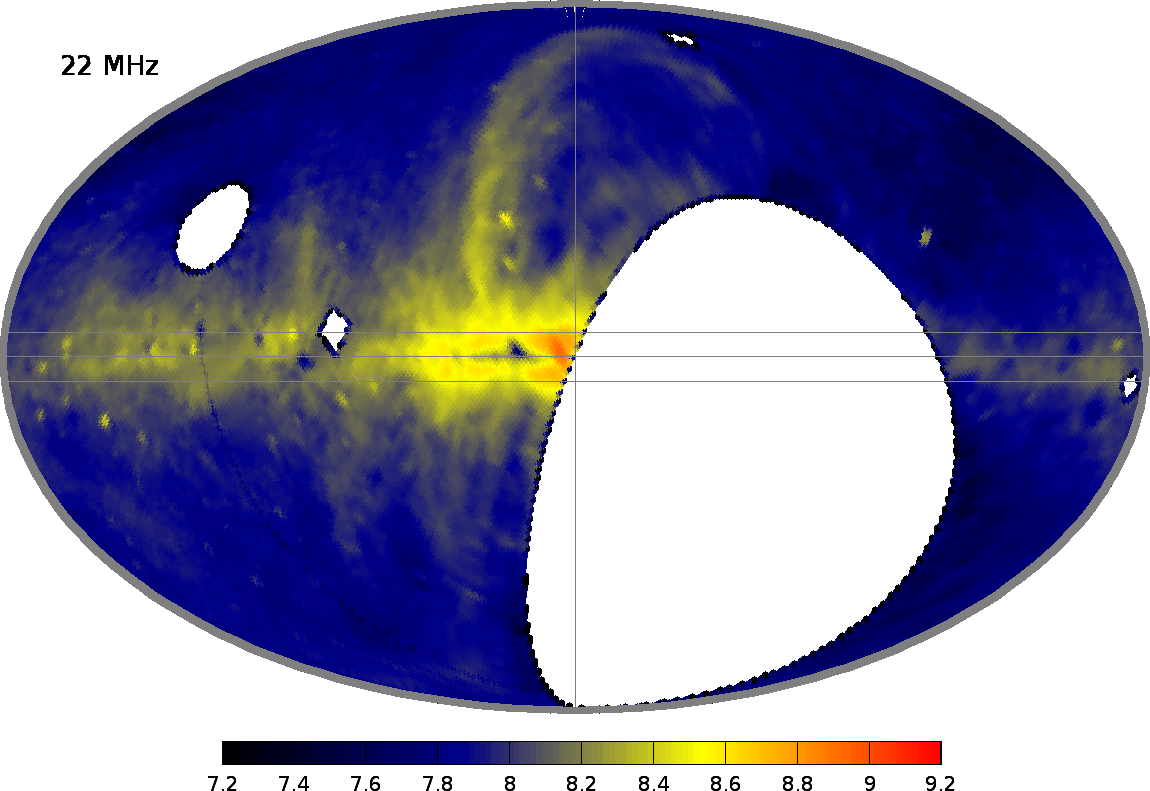}}\\[3ex]
	\resizebox{\textwidth}{!}{
	\includegraphics[width=0.49\textwidth, bb = 20 20 1130 775]{skymap_45mhz.png}\hspace{2ex}
	\includegraphics[width=0.49\textwidth, bb = 20 20 1130 775]{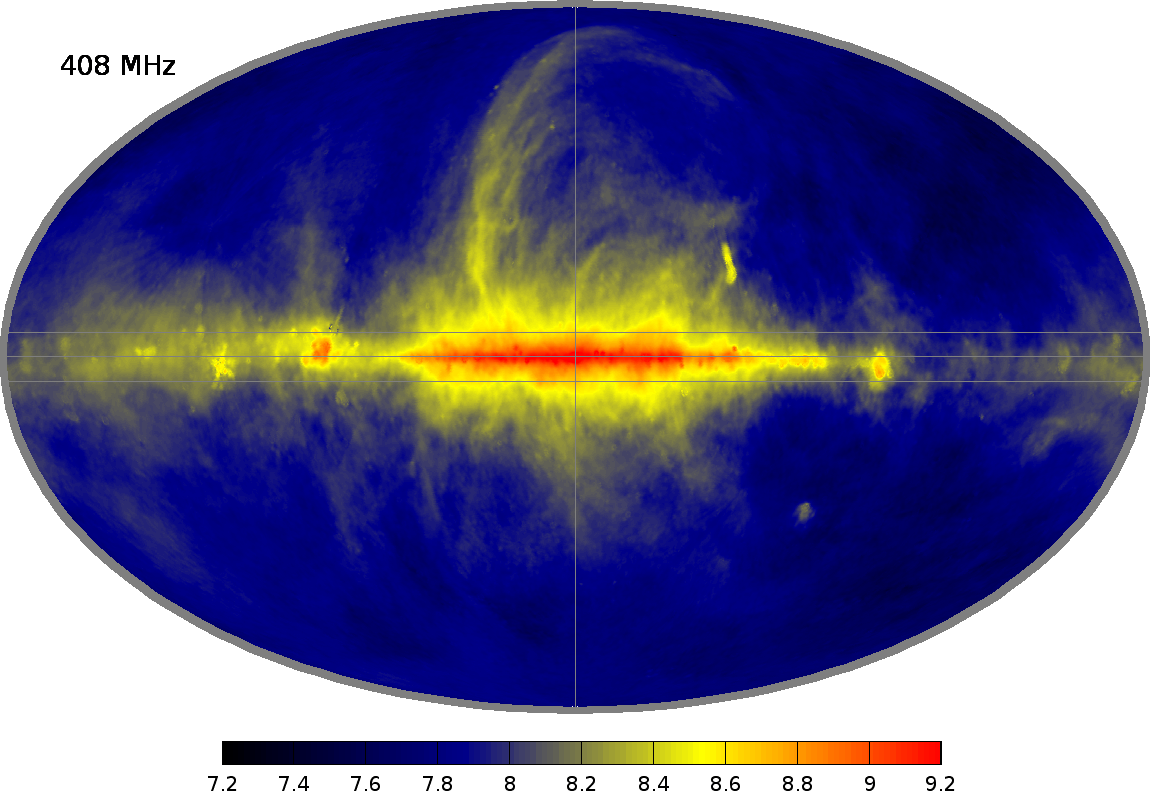}}\\[3ex]
	\resizebox{\textwidth}{!}{
	\includegraphics[width=0.49\textwidth, bb = 20 20 1130 775]{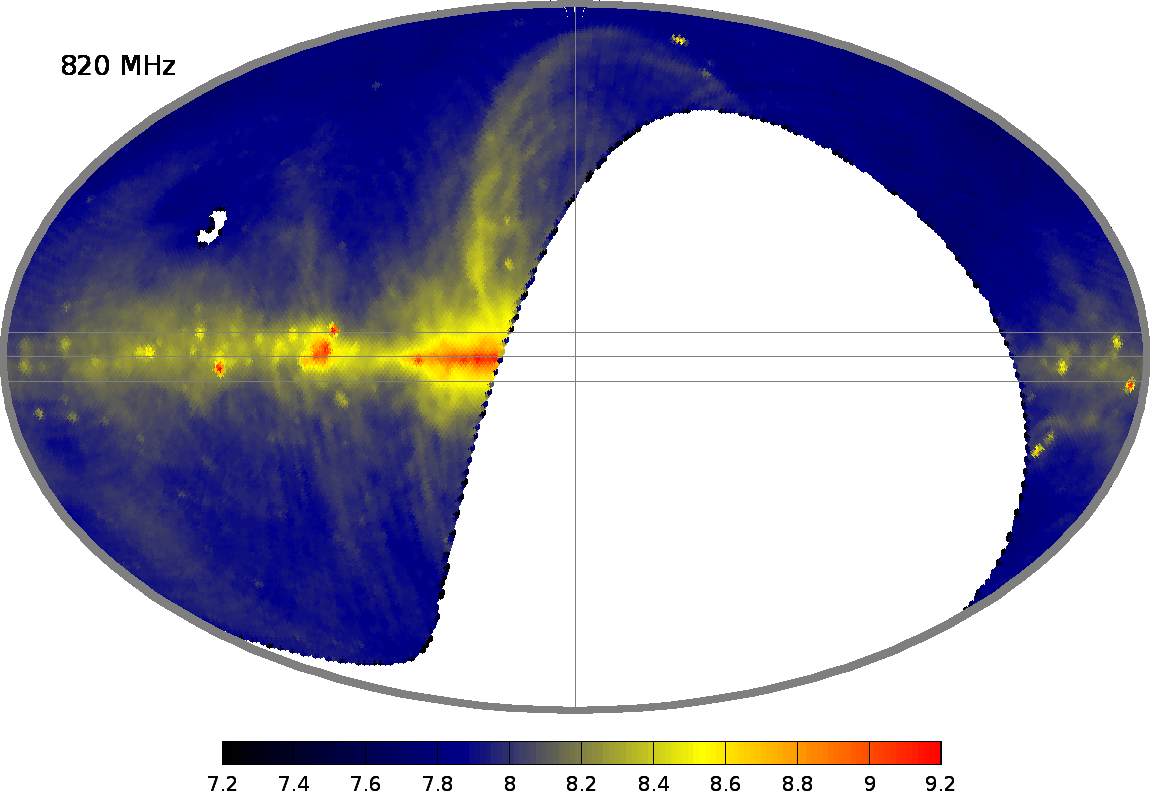}\hspace{2ex}
	\includegraphics[width=0.49\textwidth, bb = 20 20 1130 775]{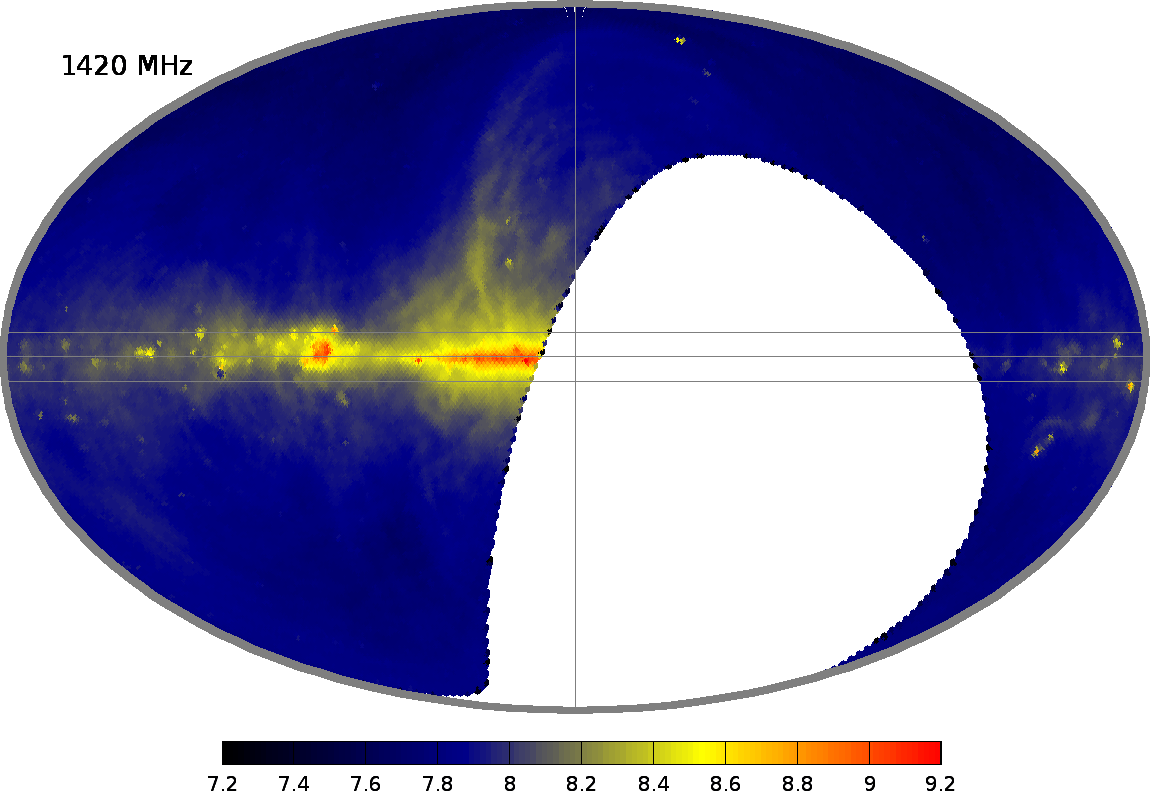}}
	\caption{Temperature skymaps (from top to bottom and left to right) at 22, 45, 408, 820, and 1420 MHz. See table~\ref{tab:surveys}. Color scale is as in figure~\ref{skymap-mu}.}
	\label{surveys}
\end{figure*}

\begin{figure*}[tb]
	\centering
	\includegraphics[width=0.69\textwidth, bb = 60 140 750 550]{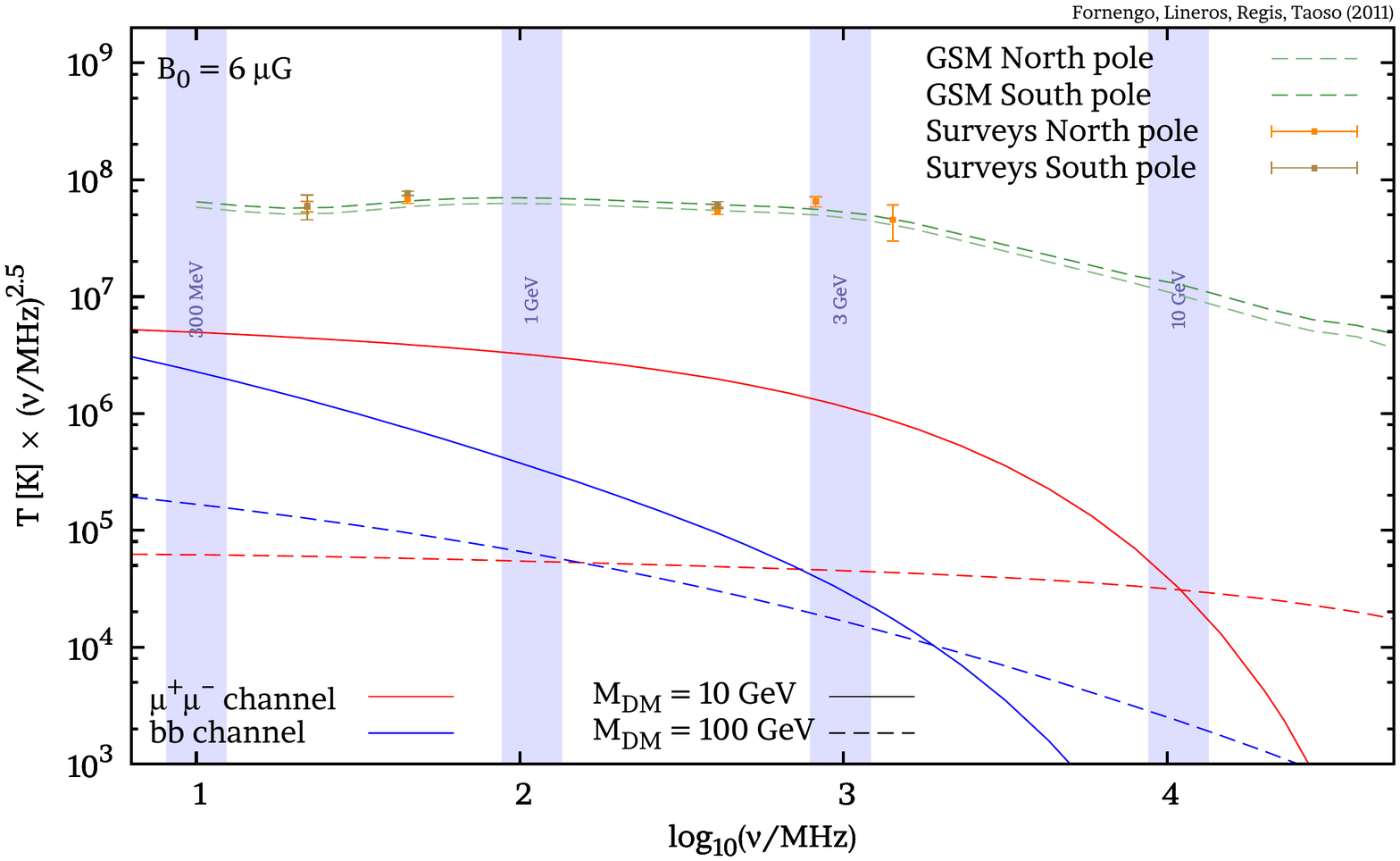}
	\caption{Temperature versus frequency calculated for the galactic poles $(b=\pm 90^{\circ})$ for $\mu^{+}\mu^{-}$ and $b\bar{b}$ annihilation channels, $M_{\rm{DM}} = 10,100~{\rm GeV}$. The DM profile and the galactic magnetic field are as in fig.~\ref{skymap-mu} and the propagation model is the MED one, table~\ref{tab:models_res}. 
The data points are the temperature at north and south galactic poles averaged in a $10^{\circ}$ circle.
Green dashed lines are linked to observations and have been obtained with the software developed in~\cite{deOliveiraCosta:2008} (see text for more details).}
	\label{freq_dependence}
\end{figure*}

From the different radio surveys available, we select those in Table~\ref{tab:surveys} and plotted
in Fig.~\ref{surveys}. They all refer to observations at $\lesssim$~GHz frequencies and have a large sky coverage.
In Fig.~\ref{freq_dependence} we show the average temperature observed at the galactic
poles at different frequencies.
The data have been averaged over a $10^{\circ}$ circle.
For illustration, we also plot the temperature as a function of the frequency obtained
using the software presented in Ref.~\cite{deOliveiraCosta:2008}, which produces skymaps
of the radio emission in a large range of frequencies interpolating between available data for the frequencies and the regions of the sky not covered by observations.

The vertical bands illustrate the average frequency of the synchrotron radiation produced by electrons of
a given energy and suggest that the the synchrotron emission produced by
($\lesssim100$~GeV) DM candidates peaks at sub--GHz frequencies.
%
These frequencies are therefore particularly suitable to search for light/intermediate DM masses. 
Moreover, both WIMP models with dominant hadronic/bosonic annihilation final state and leptophilic light WIMP models induce a synchrotron spectrum which is softer than the observed galactic one, as shown in Fig.~\ref{freq_dependence}.
One of the main motivations of this paper resides therefore in the fact that
low radio frequencies have the largest constraining power for such WIMP models.

\begin{figure*}[tb]
	\centering
	\includegraphics[width=0.69\textwidth,bb = 10 10 910 510]{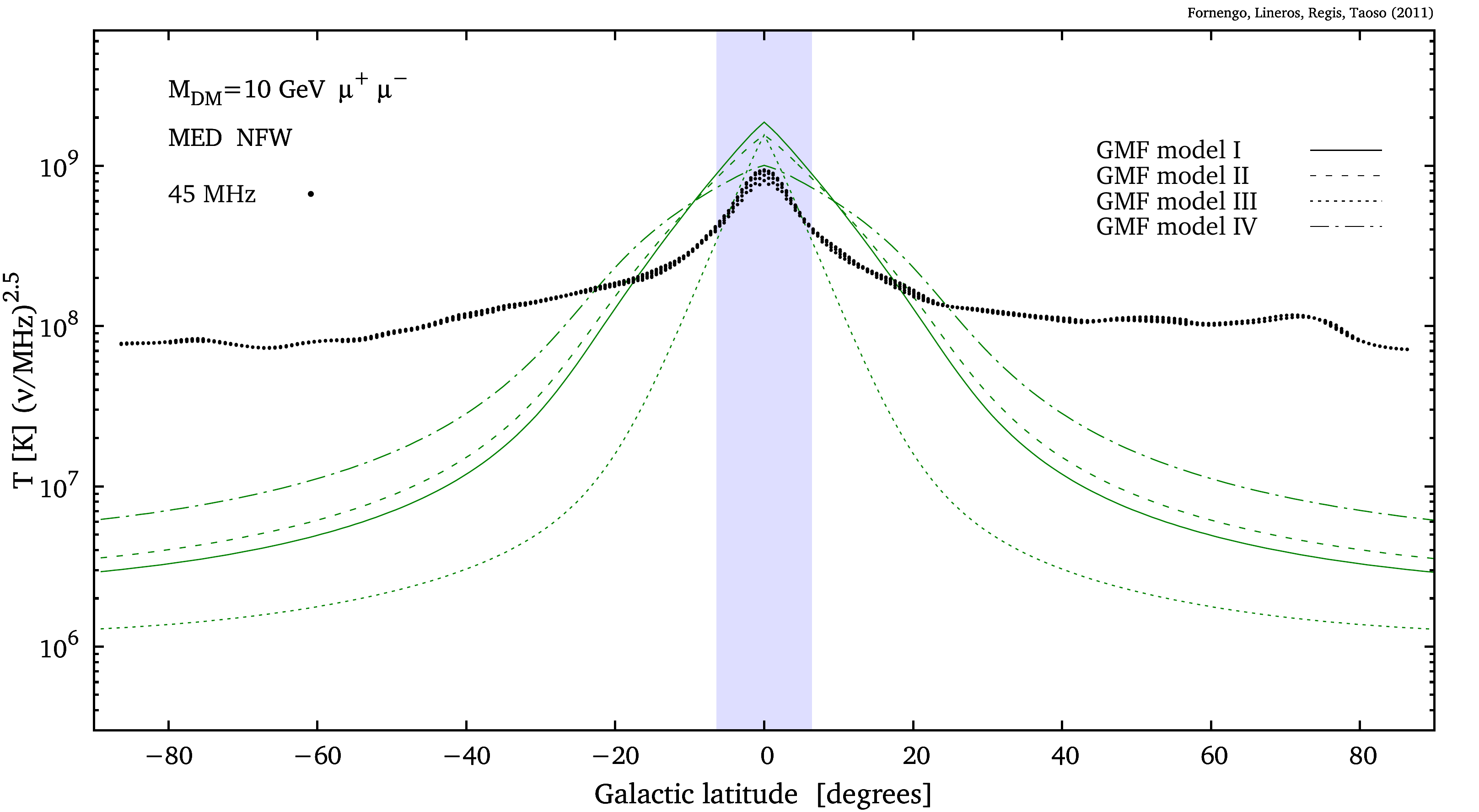}
	\includegraphics[width=0.69\textwidth,bb = 10 10 910 510]{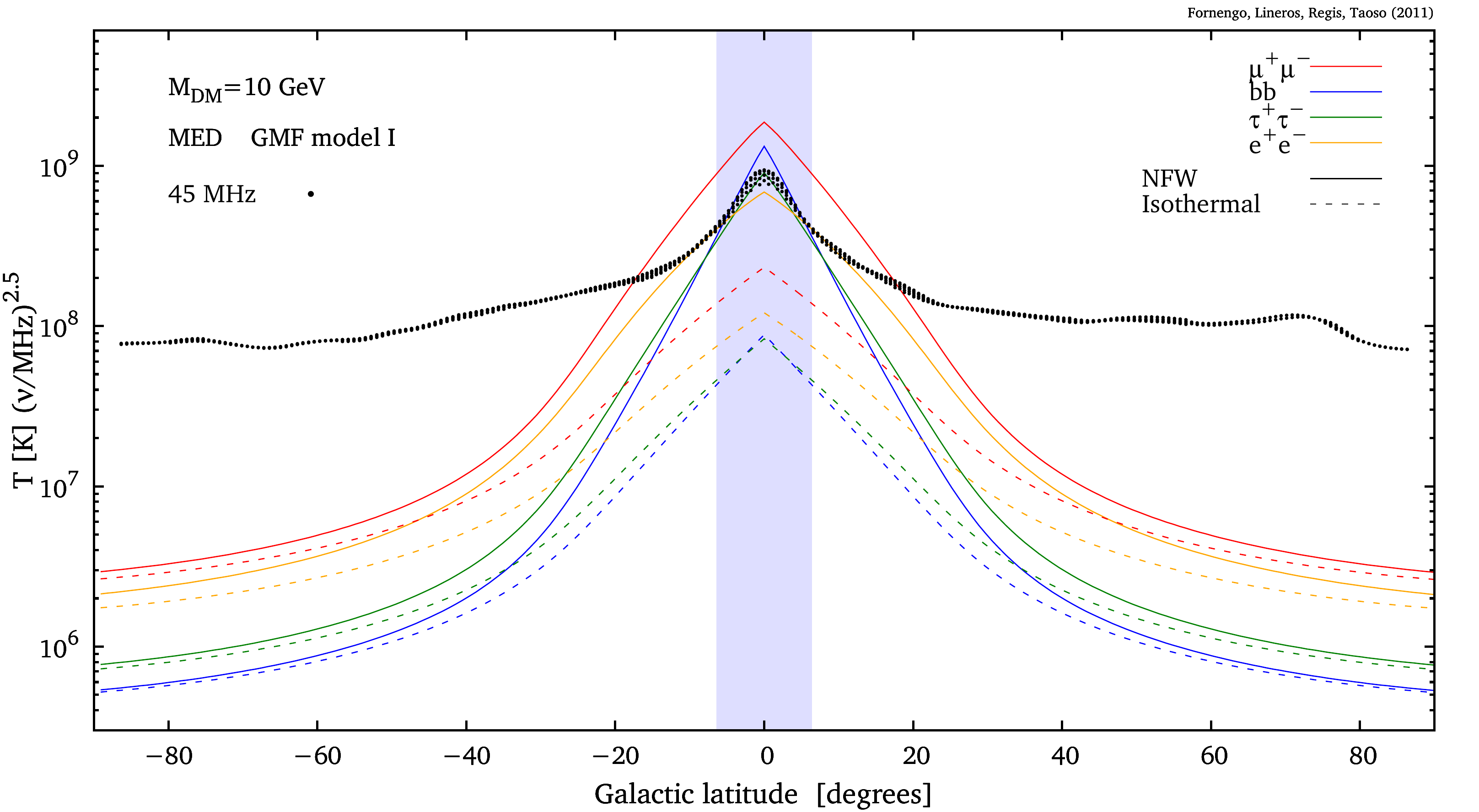}
	\includegraphics[width=0.69\textwidth,bb = 10 10 910 510]{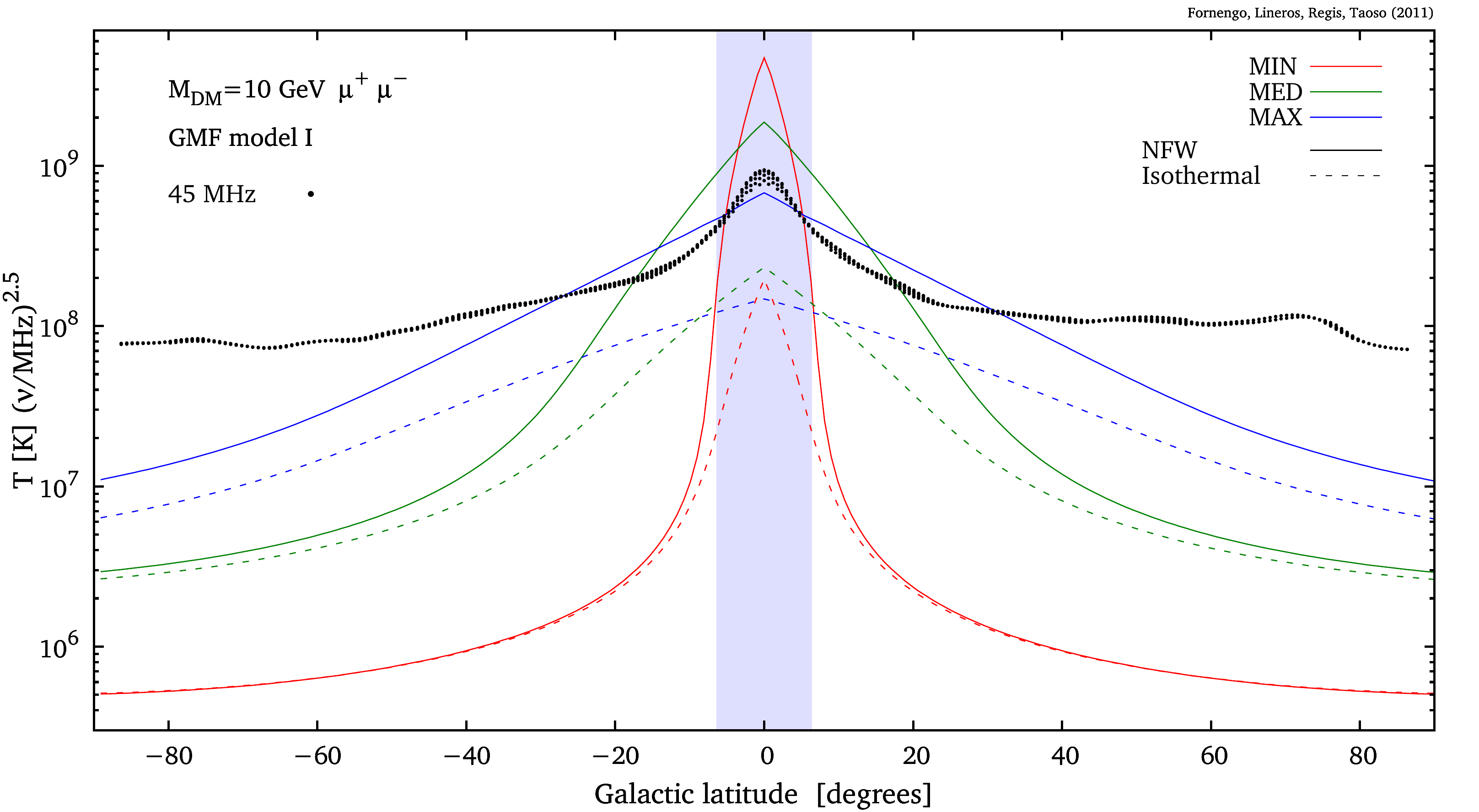}
	\caption{\label{45Mhz0_l0_strip}
Temperature versus galactic latitude at 45~MHz. 
The data corresponds to a thin strip $|l| < 3^{\circ}$, with $l$ the galactic longitude. 
Lines are predictions for DM models for $l=0^{\circ}$.
Blue band corresponds to $|b| < 10^{\circ}$ and it denotes the directions towards to the Galactic center region.
Upper--panel shows prediction for GMF models described in table~\ref{tab:mag_models}, for NFW profile, MED model, and $\mu^{+}\mu^{-}$ channel.
Middle--panel shows predictions for different annihilation channels, NFW and Isothermal profiles with GMF set to model I and propagation to the MED model.
Lower--panel shows predictions for the $\mu^{+}\mu^{-}$ channel with the all three propagation models: MIN, MED, and MAX, and with the profiles: NFW and Isothermal. }
	
\end{figure*}

The plots in Fig.~\ref{45Mhz0_l0_strip} compare observations and DM emissions at $45$~MHz for different DM models and astrophysical setups.
The observational data correspond to a thin strip crossing the galactic center and perpendicular to the galactic plane ($|l| < 3^{\circ}$).
We notice that the emission is more extended for propagation models with larger $L_z$.
This means that for the MED and MAX models it is possible to search
for the DM signal outside the inner galactic center region $(|b| > 10^{\circ})$.
Inside this region, depicted by the blue band, the astrophysical uncertainties on the propagation model, the DM distribution, and the contamination from the astrophysical background become more important.
%
As expected, for the cored isothermal profile, the emission is strongly reduced at small galactic latitudes and longitudes. 
The scaling of the signal with respect to the magnetic field strength is $\propto B^{-2}$.
In general, the emission does not dramatically change for different GMF spatial profiles, unless one considers a large mismatch between the size of the diffusion box and the scale at which the magnetic field becomes suppressed, as it happens for the GMF IV together with the MED and MAX propagation models.

While the observed radio emission is dominated by the astrophysical background,
Fig.~\ref{45Mhz0_l0_strip} suggests that DM  could substantially contribute to
the radio flux, especially close to the galactic center region.
Certainly a possible DM detection is challenged by the large uncertainties
which affect the determination of the background.
In the next section, in order to be conservative 
we just use the present observational data to constrain DM models,
without attempting any background subtraction,
and we comment on the possibility to detect DM signals with radio observations
in Section~\ref{prospects}.

\section{Constraints}
\label{constraints}

\begin{figure*}[t]
	\centering
	\resizebox{\textwidth}{!}{
	\includegraphics[width=0.48\textwidth, bb = 10 20 910 510]{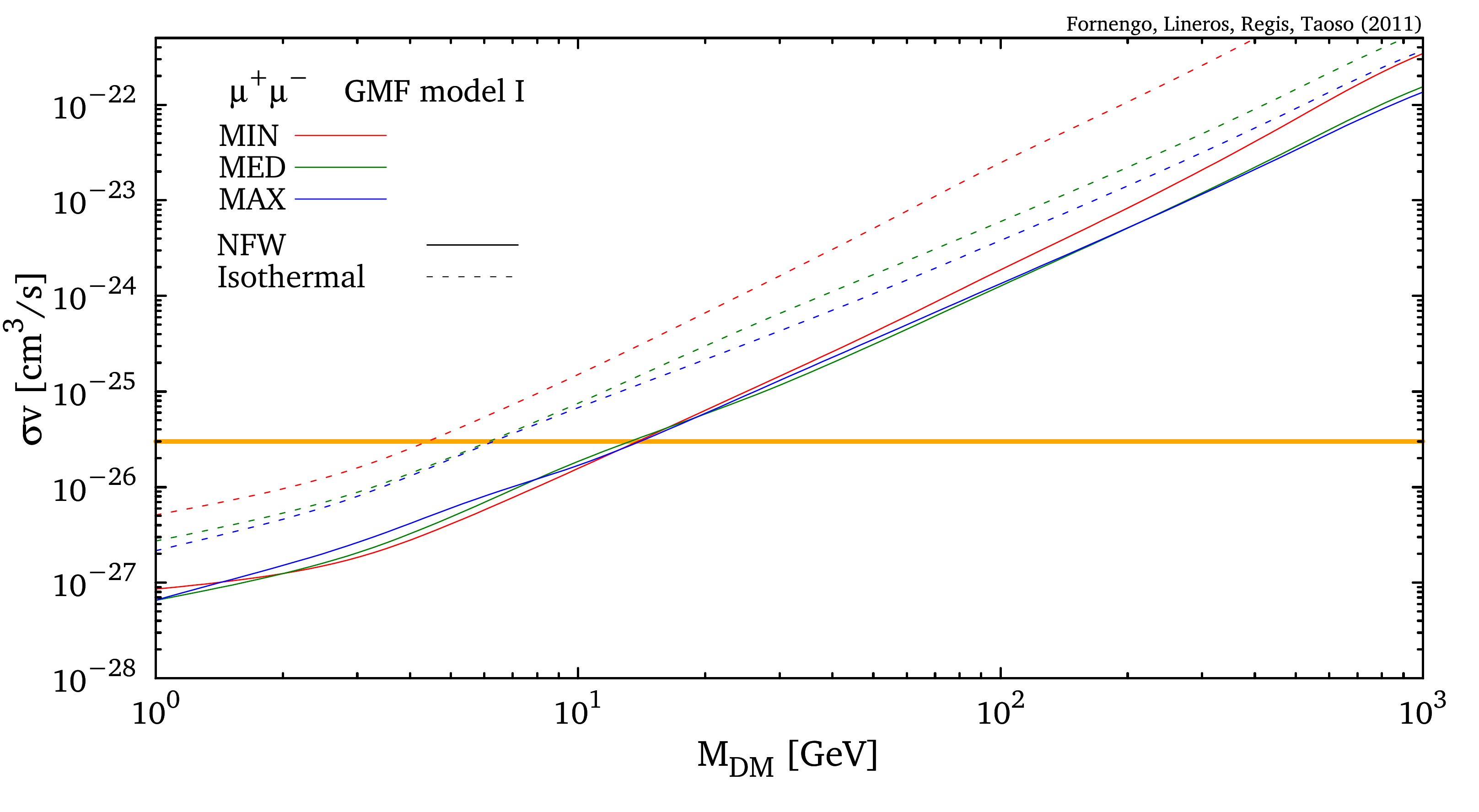}
	\includegraphics[width=0.48\textwidth, bb = 10 20 910 510]{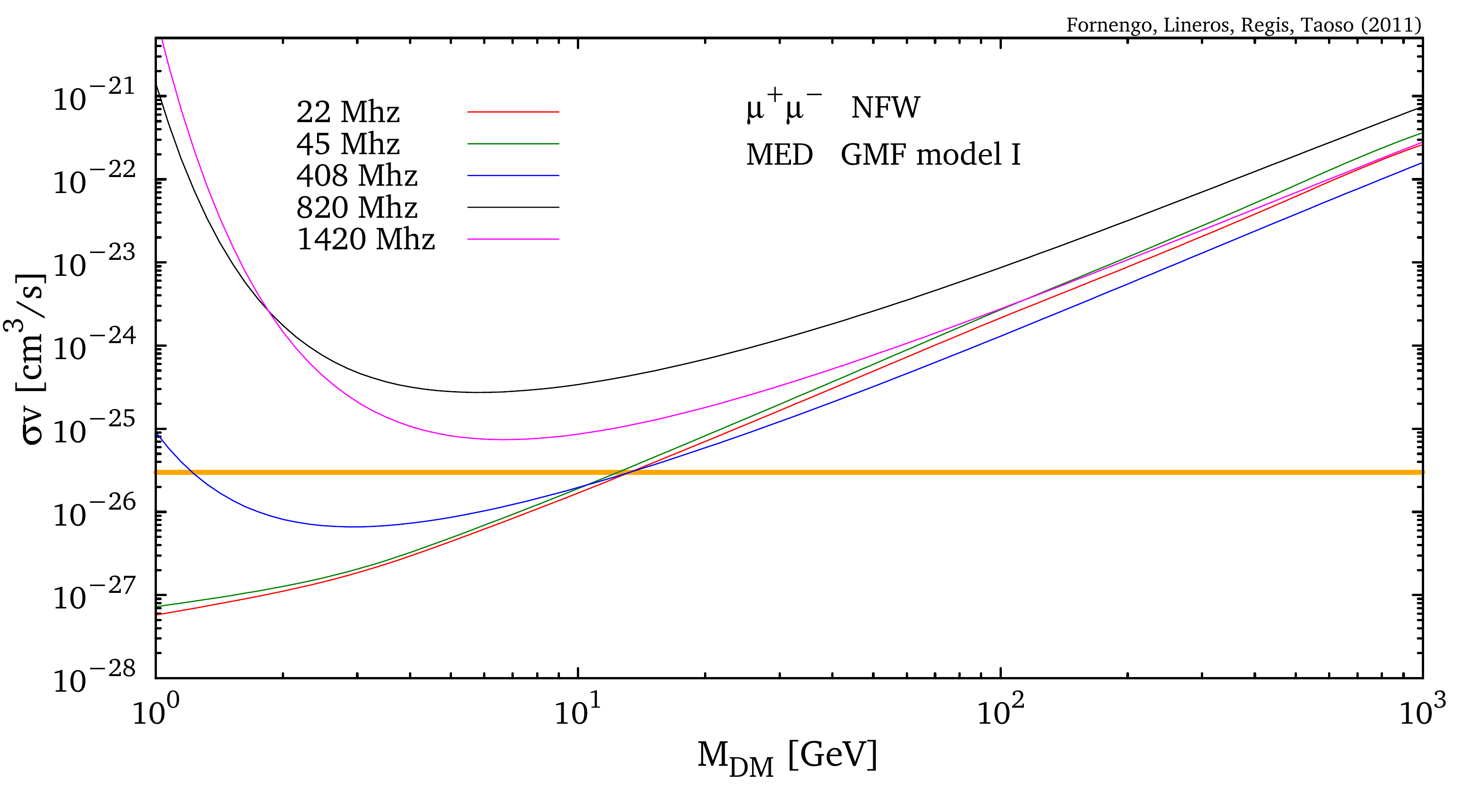}}
	\resizebox{\textwidth}{!}{
	\includegraphics[width=0.48\textwidth, bb = 10 20 910 510]{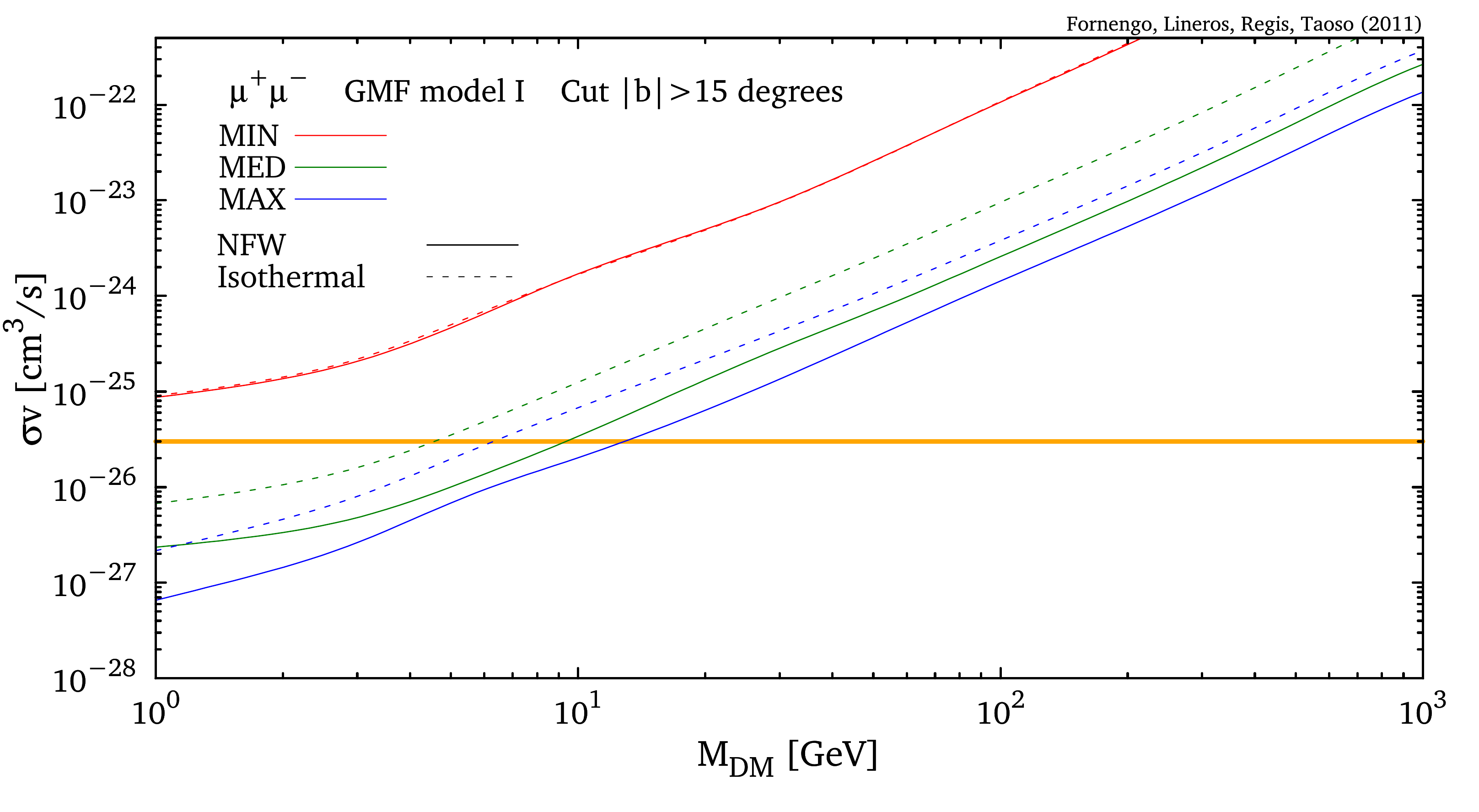}
	\includegraphics[width=0.48\textwidth, bb = 10 20 910 510]{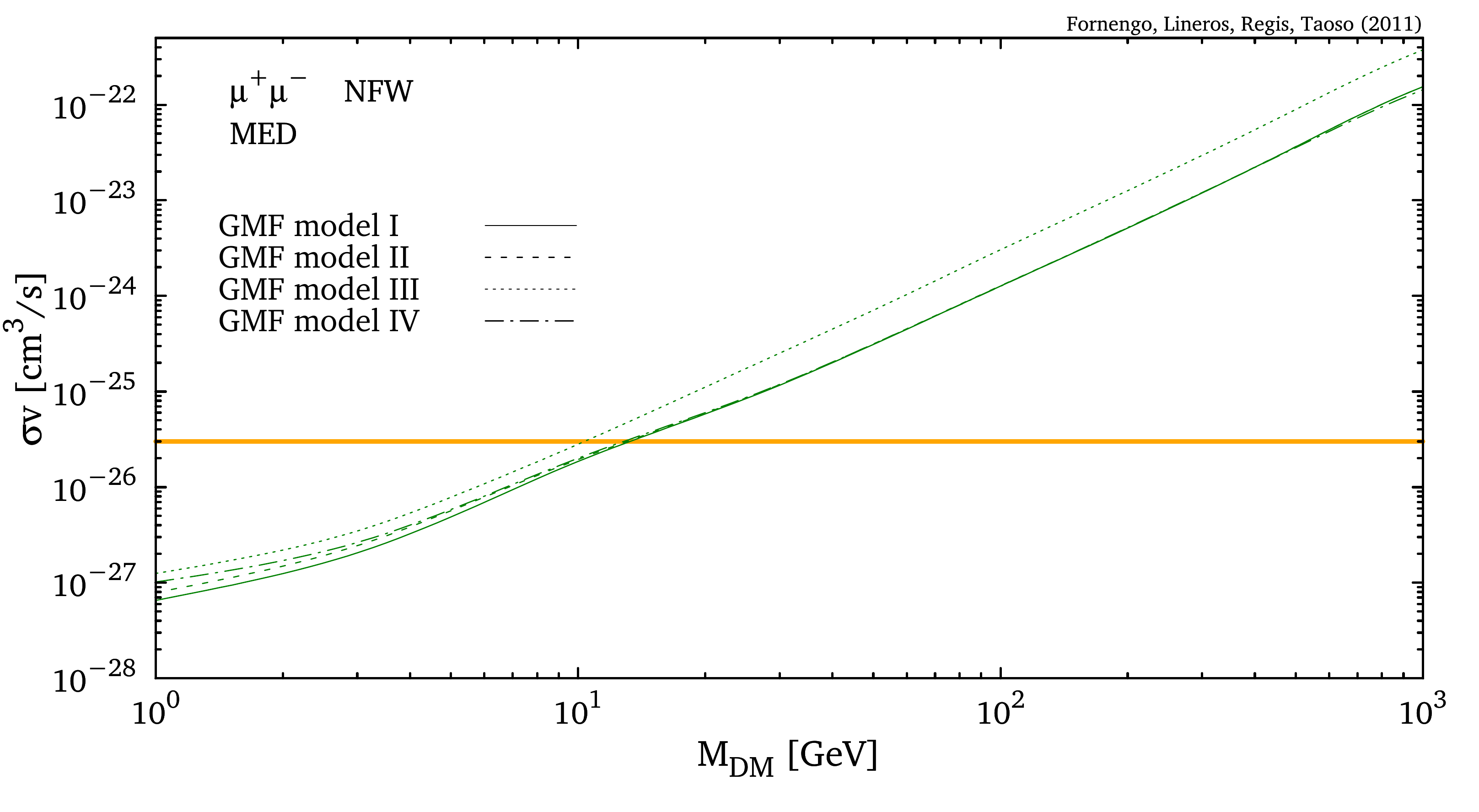}}

	\caption{	\label{sigmav_mu}
Upper bounds on the annihilation cross section $(\sigma v)$ as a function of the DM mass for the muonic channel.
Different panels are for different astrophysical setups. 
Upper--left panel shows the constraints inferred with all survey using the whole sky.
The constraints for NFW profile are more stringent than the Isothermal case.
Upper--right panel shows the constraints inferred individually from each of the surveys considered in the analysis.
In the lower--left panel, the bounds are obtained excluding the strip along the galactic plane $|b|<15^{\circ}$.
This means that in the grid of appendix~\ref{app:grid} we have excluded three lines of patches around the galactic plane.
Constraints inferred for different GMF models are presented in the lower--right panel.}
\end{figure*}

\begin{figure*}[t]
	\centering
	\resizebox{\textwidth}{!}{
	\includegraphics[width=0.48\textwidth,bb = 10 20 910 510]{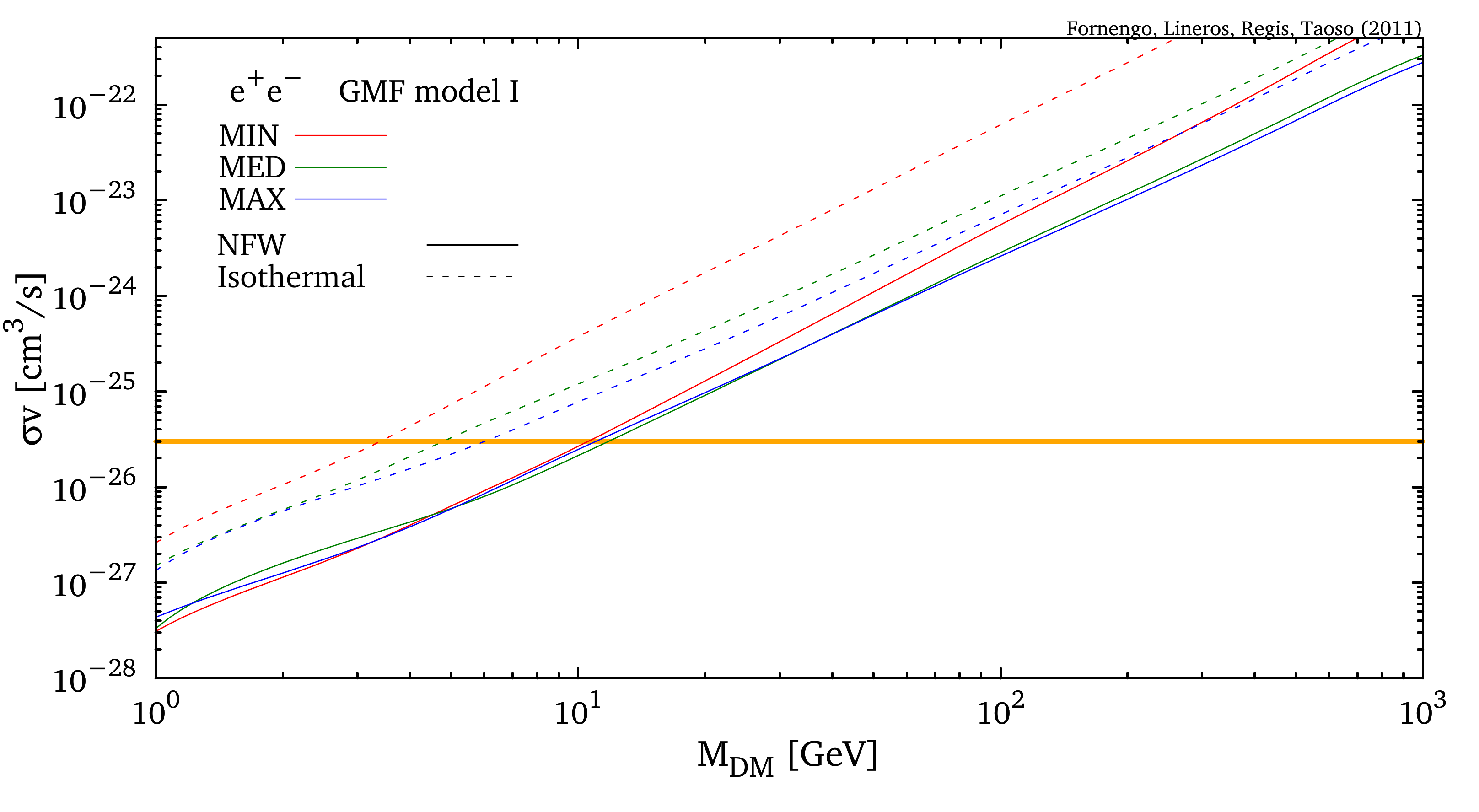}
	\includegraphics[width=0.48\textwidth,bb = 10 20 910 510]{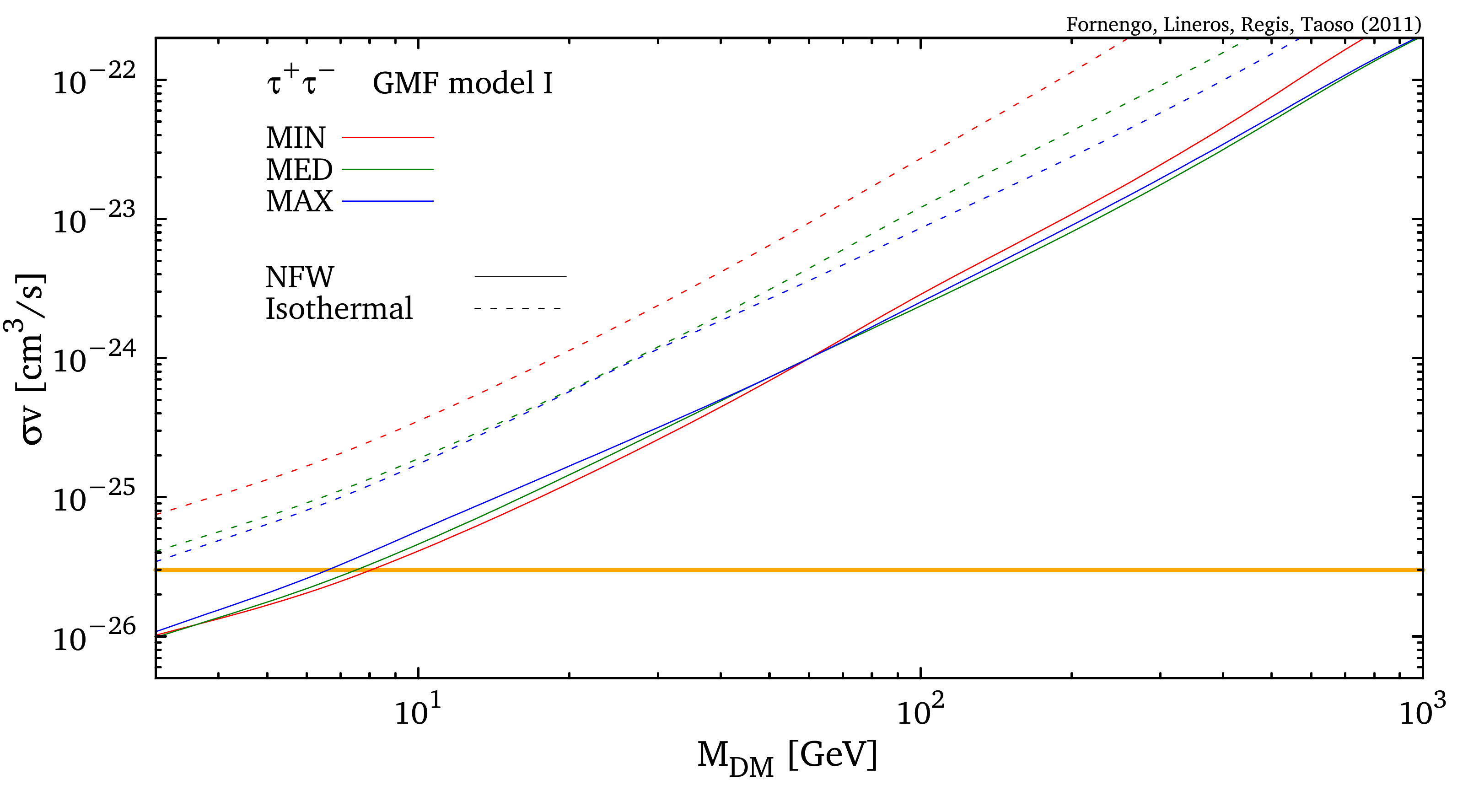}}
	\resizebox{\textwidth}{!}{
	\includegraphics[width=0.48\textwidth,bb = 10 20 910 510]{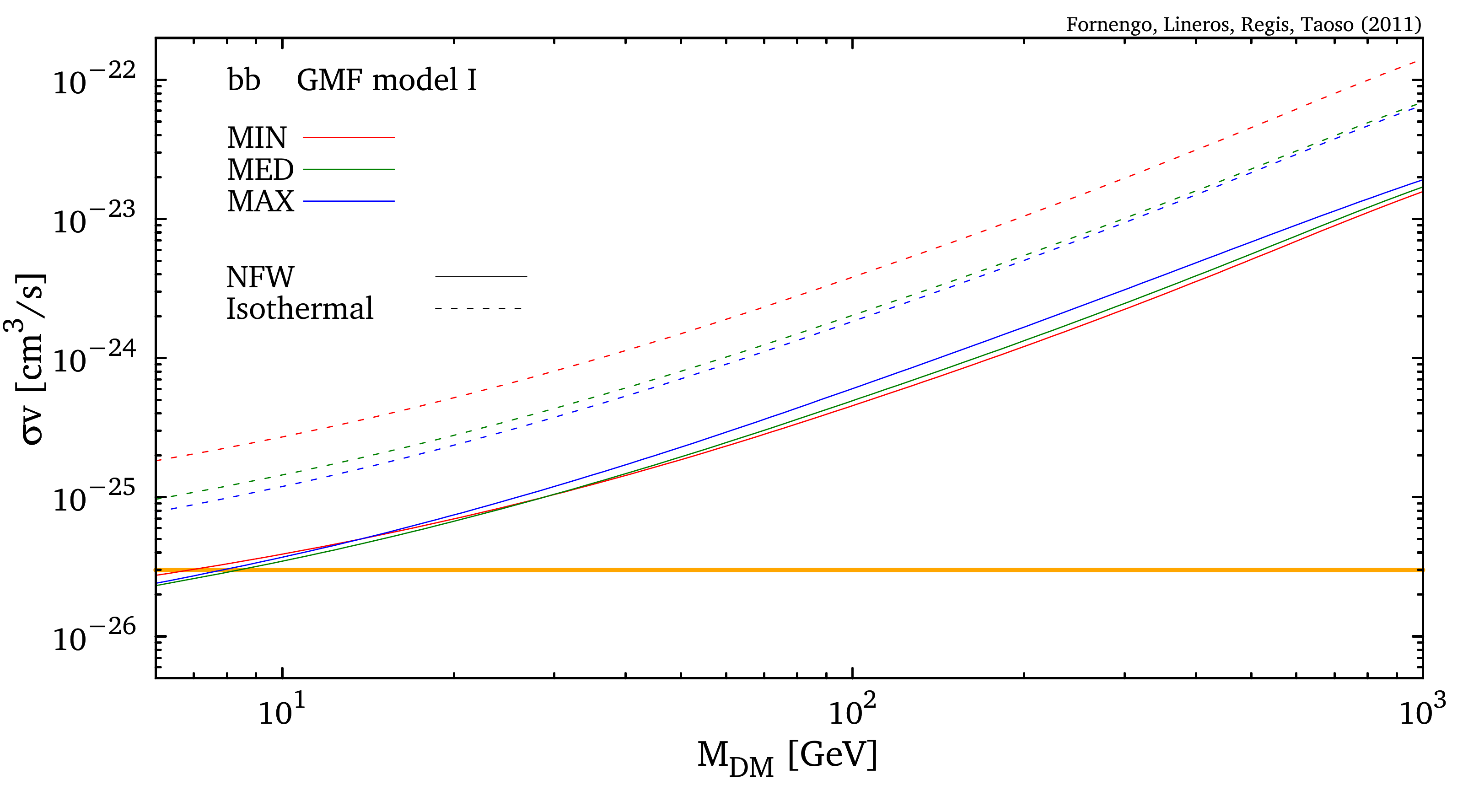}
	\includegraphics[width=0.48\textwidth,bb = 10 20 910 510]{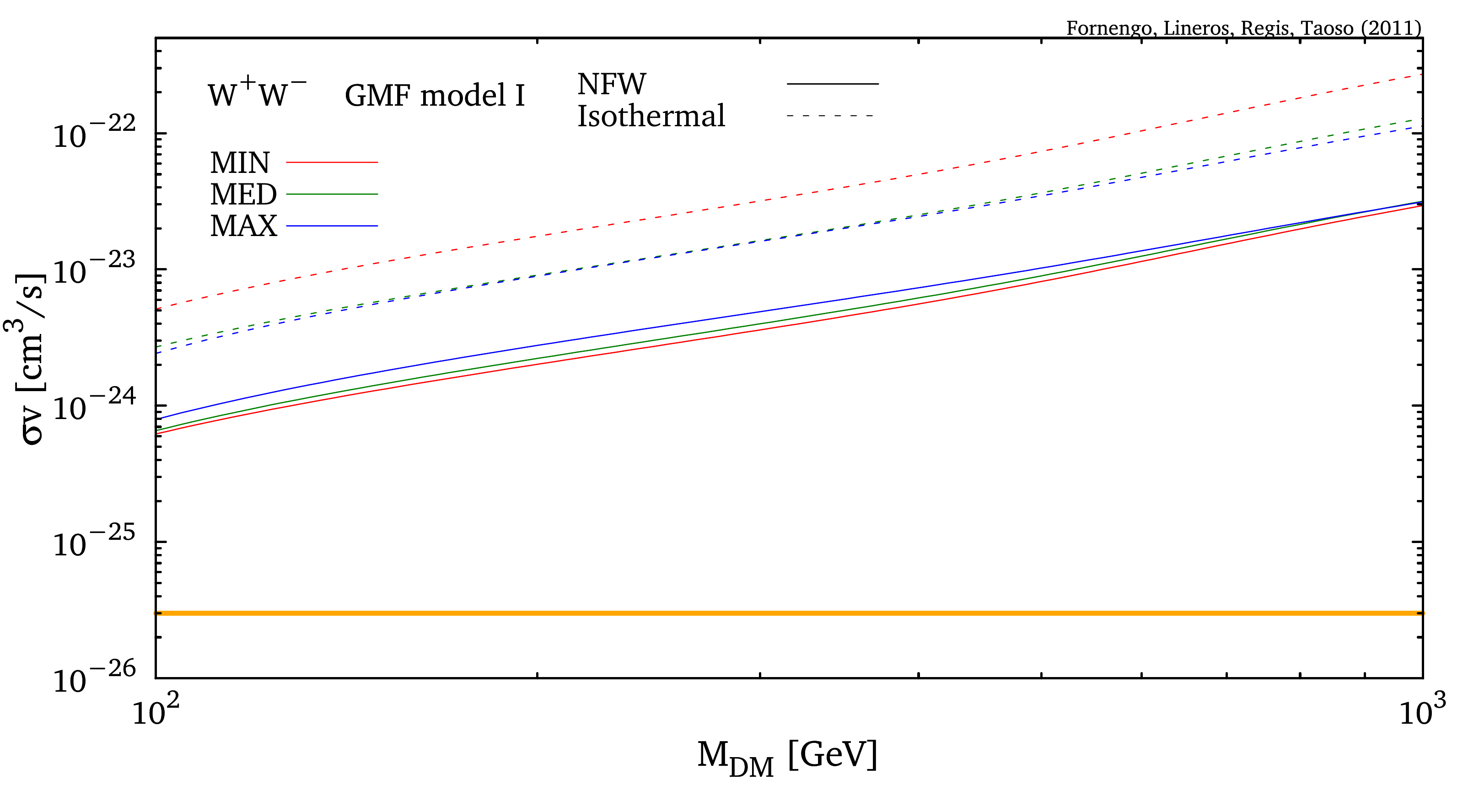}}

	\caption{Upper bounds on the annihilation cross section $(\sigma v)$ as a function of the DM mass for different annihilations channels: $e^{+}e^{-}$ (Upper--left), $\tau^{+}\tau^{-}$ (Upper--right), $b\bar{b}$ (Lower--left), and $W^{+}W^{-}$ (Lower--right).  
In most of the cases, the thermal cross section value, orange solid line, is reached at $M_{\rm DM} \sim 10 ~{\rm GeV}$.
}
	\label{sigmav_ch}
\end{figure*}

The bound on the dark matter parameter space are conservatively inferred requiring that
dark matter synchrotron emission does not overshoot the data, without taking into account the
contribution of the astrophysical background. 
Our procedure is the following: the sky is divided in 232 patches obtained as explained in Appendix~\ref{app:grid}.
Then we compute the average DM radio emission ($T_{\rm{DM}}$) and average observed radio flux ($T_{\rm{obs}}$) in all the patches which are covered by the data (note that most of the surveys do not cover all the sky, see Fig.~\ref{surveys}).
The constraint inferred from each patch is obtained requiring that: 
\begin{equation*}
	T_{\rm{DM}} \leq T_{\rm{obs}} + 3 \sigma
\end{equation*}
where $\sigma$ is the rms temperature noise that we take as in Table~\ref{tab:surveys}.
We repeat the calculation for all the frequencies and then we set a bound on
the DM annihilation cross section $(\sigma v )$ by taking the most constraining patch.
The results are summarized in Figs.~\ref{sigmav_mu} and~\ref{sigmav_ch}.

The bounds do not dramatically vary for different annihilation channels.
We find that models with DM masses $M_{\rm{DM}} \lesssim 10$~GeV and thermal value of the annihilation cross section $(\sigma v) = 3 \times 10^{-26}$~cm$^3$~s$^{-1}$ are strongly constrained in the case of the NFW profile. 
We also notice that the constraints drastically weaken for the isothermal profile, which presents a much lower DM density in a large region around the Galactic center. 
This effect has also been shown in Fig.~\ref{45Mhz0_l0_strip}.
Still, for this profile and for the $\mu^{+}\mu^{-}$ and $e^{+}e^{-}$ annihilation channels, thermal values of $(\sigma v)$ are excluded for $M_{\rm{DM}} \lesssim 4-6$~GeV.

Despite the morphology of the emission is quite different for the three propagation models, the derived bounds are instead similar (except when cutting away $|b|<15^{\circ}$), as it is shown in Fig.~\ref{sigmav_mu}, where the case of DM annihilations into muons is considered.
This is because in the most constraining patches (i.e., low latitudes) the average DM emission is similar for the three cases.

Similar conclusions can be drawn about the GMF model.
Indeed, changing the magnetic field profiles does not dramatically alter the constraints, since at low latitudes the different models considered for the GMF are similar. The example shown in~Fig.~\ref{sigmav_mu} is for the MED model. 
Similar results are obtained for the MAX setup while in the MIN case bounds are weaker, but within a factor of less than 1.5.

We also study the constraints inferred individually from each survey (upper--right panel of Fig.~\ref{sigmav_mu} ).
As discussed in Sec.~\ref{comparison}, the lowest frequencies are more constraining for low
DM masses $(M_{\rm{DM}} \lesssim 10 {\rm~GeV})$ while $\mathcal{O}(\rm{GHz})$ frequencies becomes relevant for heavier DM candidates. 
Let us remark that the constraining power of a single survey also depends on the fraction of sky covered and on the sensitivity of the map,
as it is shown in Fig.~\ref{sigmav_mu} where the 820~MHz survey provides worse constraints than the 1420~MHz one.\\

As commented in Sec.~\ref{sec:cosmicrays},  additional energy losses than those we have considered here might become relevant
in the galactic plane. We estimate that this effects might reduce our predictions on the synchrotron fluxes in the region $|b| \lesssim 1^{\circ}$,
so their impact on the bounds would be rather small, since they are derived considering patches of the sky with significantly larger angular sizes.
Still, we decide to compute the bounds by cutting a large region around the galactic plane by imposing
$|b|<15^{\circ}$ (lower--left panel in Fig.~\ref{sigmav_mu}), which in our case is equivalent to remove 3 strips of patches. 
In general, with this conservative choice, we reduce the uncertainties on the propagation related to possible unaccounted astrophysical
effects occurring on the galactic disk.
Moreover, the fact that the radio astrophysical background and its uncertainties are maximal in the galactic plane,
complicates the searches of DM signals in this region.
Therefore, by avoiding the region $|b|<15^{\circ},$ we study the reach of current radio surveys in
the DM parameter space focusing on "cleaner" targets of observations.
%
%
Interestingly, we find that the bounds are almost unchanged for the MAX model and only slightly affected for the MED one. 
For the MIN model, they are more altered since the emission is more concentrated in the Galactic center region.
These results are consistent with what discussed in the Sec.~\ref{comparison}.

\section{Prospects for detection}
\label{prospects}

So far our goal has been to derive robust and conservative bounds on WIMP parameter space.
We now analyze the possibility that the data already contain a significant DM contribution, and discuss
a method to single it out. 

As for the constraints, our focus is on WIMP models inducing a synchrotron emission with a spectrum softer than
the astrophysical one (see Fig.~\ref{freq_dependence}) with the latter mainly coming from cosmic--rays and showing a spectrum roughly proportional to $\nu^{-2.5}$.
As it is clear from Fig.~\ref{45Mhz0_l0_strip}, a possible DM excess would be maximal in the central region of the Galaxy (within 20--30 degrees), and should appear as a spherical feature related to the (approximately) spherical shape of the DM profile. This is opposite to what is expected for the CR emission, which typically shows a ``disky'' shape, following from the confinement of CR sources to the stellar disc.
Therefore, we perform a search where the signal to background ratio is expected to be larger, i.e. in maps at low frequencies and in the inner part of the Milky--Way.
We choose the 45 MHz map~\cite{Guzman2010} where the central region is better sampled, but analogous analysis can be done with the 22 MHz map~\cite{1999A&AS..137....7R}.

\begin{figure*}[t]
	\centering
\resizebox{0.9\textwidth}{!}{
	\includegraphics[width=0.28\textwidth, bb = 0 20 725 421]{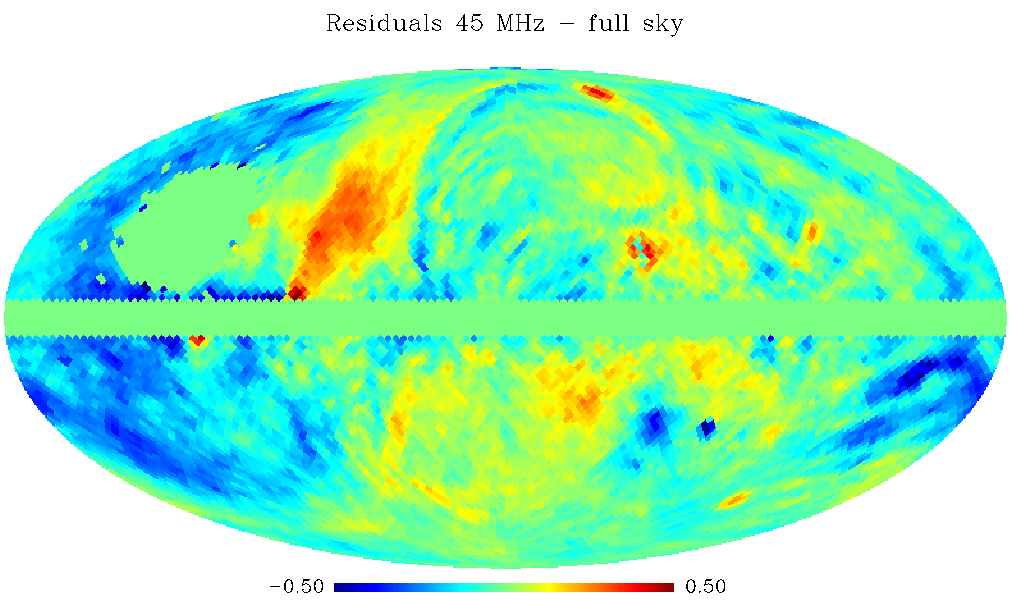}
        \hspace{3ex}
	\includegraphics[width=0.14\textwidth, bb = 0 40 425 510]{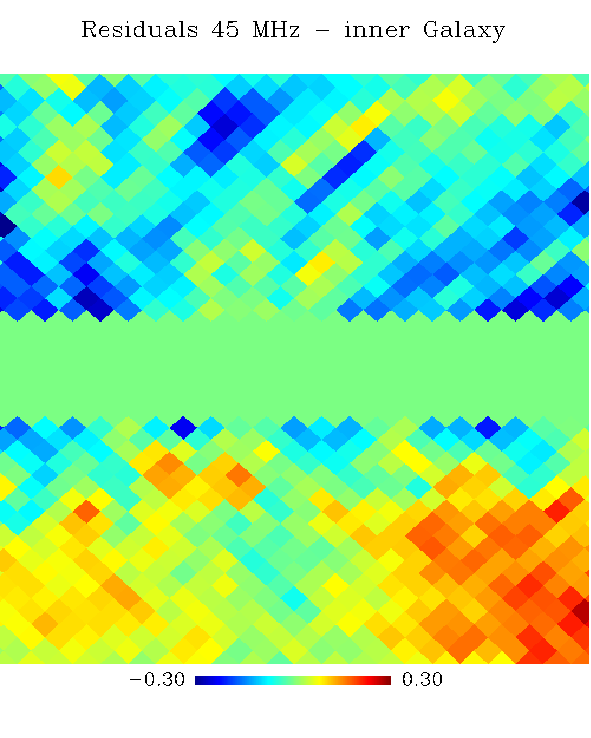}
}

	\caption{Residuals $R_{i}=(T^{\rm est}_{i}-T^{\rm obs}_{i})/T^{\rm obs}_{i}$ at 45 MHz for the full-sky (left) and in the inner $30^\circ\times30^\circ$ of the Galaxy (right). Haslam et al. map~\cite{1982A&AS...47....1H} at 408 MHz is used as a template to subtract galactic synchrotron emission. See text for details.}
	\label{fig:resid}
\end{figure*}

We consider the Haslam et al. map~\cite{1982A&AS...47....1H} at 408 MHz as a template for galactic synchrotron emission. 
This is commonly done also in CMB studies since the Haslam map is the radio full--sky map with the best angular resolution and sensitivity, at a frequency where the emission is thought to be dominated by synchrotron radiation.
We estimate the emission at 45 MHz in each angular pixel as $T^{\rm est}_{i}=T_{408,i}\cdot(45 {\rm MHz}/408 {\rm MHz})^{\alpha}+T^0$, where $i$ is the pixel index, and the 408 MHz map is smoothed down at the angular resolution of the 45 MHz map \cite{Guzman2010}.
$\alpha$ and $T^0$ are derived from the best--fit of the observed temperature $T^{\rm obs}$ in the map of Ref. \cite{Guzman2010}, by minimizing the $\chi$--square function $\chi^2=\sum_i (T^{\rm est}_{i}-T^{\rm obs}_{i})^2/\sigma^2$, where $\sigma$ is the noise--level reported in Table~\ref{tab:surveys}, and in the sum we include all available pixels in the map except for the disc ($|b|<5^\circ$). We include $T^0$ in the fit to account for a possibly different spectral index in the extragalactic emission, or for possible experimental issues associated to the absolute normalization of the flux. However, it is nearly irrelevant since its best--fit value comes out to be at the level of noise. 
We found $\alpha=-2.56$, confirming our expectation, and suggesting that this search technique can be indeed useful for DM candidates inducing a synchrotron spectrum softer than $\sim \nu^{-2.5}$.
The residuals $(T^{\rm est}_{i}-T^{\rm obs}_{i})/T^{\rm obs}_{i}$ are shown in Fig.~\ref{fig:resid} for the central $30^\circ\times30^\circ$ box, and in the full-sky.
Note that this technique is similar to the one employed for the ``WMAP haze''~\cite{Hooper:2007kb,Dobler:2007wv,Cumberbatch:2009ji}, namely a possible foreground excess found in the inner Galaxy at microwave frequencies and which can be interpreted in terms of WIMP annihilations/decays.
Contrary to the possibility investigated in this Section, the WMAP haze involves an excess which is harder than the rest of the galactic synchrotron emission. Therefore, in that case, the focus is on high frequencies and different DM candidates.

No evidence for a bright and spherical spot is found in Fig.~\ref{fig:resid}, with the residuals fluctuating between $\pm 30\%$ in the central region. 
Therefore we found no hints for a DM signal in current data. Similar conclusion can be drawn considering the 22 MHz map~\cite{1999A&AS..137....7R}.
However, as clear from Fig.~\ref{fig:resid}, our estimates of the temperature are not matching well with measured ones, with large residual fluctuations which are left after subtraction all over the sky.
From a physical point of view, even assuming that the emission is completely due to synchrotron radiation, one can expect a spatial variation in the spectral index (which is assumed to be constant in our template technique), associated to variations of galactic magnetic properties or energy losses across the Galaxy (which in turn affect cosmic--ray spectrum and emission). 
This should be correlated to the galactic disc, while fluctuations in Fig.~\ref{fig:resid} are not. Therefore they should have a different origin.
Most of the brightest spots in the full-sky image can be ascribed to known emissions (LoopIII, Cignus region, Cen A, Virgo, North Polar Spur, and Gum Nebula to mention the most relevant) other from the diffuse galactic synchrotron radiation.
However, there is still some large residual left (see, e.g., bottom-right corner in the central region map).

On the other hand, both the derived spectral index and residuals are in fair agreement with results found in Ref.~\cite{deOliveiraCosta:2008}.
Adopting a principal component analysis, Ref.~\cite{deOliveiraCosta:2008} estimated that the first principal component at low radio frequency can be in first approximation identified with a synchrotron component with a spectral index $\alpha\sim-2.5$ and accounting for $\sim70\%$ of the total emission, while the remaining $\sim30\%$ arises from other components.

The large residuals can be therefore due either to other physical contributions (i.e., 
free--free emission, spinning dust or thermal dust), although they are expected to be largely subdominant, or to possible spurious effects due to poor angular resolution and sensitivity (e.g., one can see, even ``by eye'' from Fig.~\ref{surveys}, that the emission in the map at 45 MHz is systematically more diffuse than in the 408 MHz map).
For these reasons, our test is somewhat inconclusive, simply telling us that there is some extra-components which account for $\sim20-30\%$ of the total emission, but with properties different from the ones expected for DM.

To fully perform this test in the direction of DM searches, one would need more refined maps.
Future observations with the LOFAR telescope~\cite{lofar} have intriguingly larger possibilities of detection.
Indeed, LOFAR can go down to 15 MHz (thus enhancing the possible signal to background ratio for DM candidates discussed here) with unprecedented sensitivity, and sample the central region of the Galaxy~\cite{Kassim} with a field of view of 20--50 degree and angular resolution of 1--30 arcmin, depending on the configuration. 
This will allow to significantly improve the template fitting subtraction and can lead to significantly lower residuals, which might help to single out a DM contribution at few-percent level.

\section{Conclusions}
\label{conclusions}

We have argued that radio observations in the low-frequencies regime are particularly suitable to search
for DM annihilations in the galactic halo.
To show that, we have compared available radio surveys with the synchrotron emission produced by DM annihilations
for different DM setups and astrophysical assumptions.
We have considered DM masses in the range 1~GeV -- 1~TeV, various annihilation channels 
which imply different spectral features and, in order to bracket uncertainties in the
spatial profile, two DM density distributions: the NFW and the isothermal profiles.
Uncertainties related to the propagation of cosmic rays are taken into account considering three benchmarks propagation models, the so--called MIN, MED and MAX models.
The morphology of the corresponding synchrotron emissions are quite different, mostly because
these models have rather different sizes of the diffusion box.
Indeed, larger values of the half--height of the propagation zone $L_z$ produce more extended emissions.
We have then explored different shapes of the GMF and determined their impact on the morphology of the synchrotron emission. 

Comparing the DM synchrotron fluxes with low--frequencies radio survey
we have computed upper bounds on the DM annihilation cross section $( \sigma v )$.
Interestingly, we found that for a NFW profile a ``thermal'' cross section is strongly constrained for $M_{\rm DM} \lesssim 10~{\rm GeV}.$
The constraints we have obtained are similar or stronger than those obtained with other indirect DM searches,
like gamma-ray observations of dwarf-galaxies~\cite{Ackermann:2011}, galactic center region~\cite{Hooper:2011ti},
galaxy--clusters~\cite{Ackermann:2010rg} and isotropic diffuse background~\cite{Abdo:2010dk,Abazajian:2010sq}.
The bounds mildly depend on the propagation model and GMF profile, while they are significantly relaxed
for shallow DM distributions, like for the isothermal profile.

In the last part of the paper, we have tried to single out the presence of a possible
DM component in the present data. We considered the $408$ MHz map as a spatial template
to estimate the galactic synchrotron emission in the low frequency maps.
This method allows to search for the presence of a DM-induced component softer than the
astrophysical galactic synchrotron emission, and
the DM signal would appear as an excess in the central region with an approximately spherical shape.
We have concluded that present data does not support any evidence for the presence of this additional synchrotron component.
However, we expect that future radio surveys, in particular with the LOFAR telescope~\cite{lofar,Kassim},
will improve coverage, angular resolution, and sensitivity in low frequency radio maps, which may allow to disentangle a faint DM contribution.

{\acknowledgments
RL and MT are supported by the EC contract UNILHC PITN--GA--2009--237920, by the Spanish grants FPA2008--00319, FPA2011--22975, 
MultiDark CSD2009--00064 (MICINN) and PROMETEO/2009/091 (Generalitat Valenciana).
NF and MR acknowledge research grants funded jointly by Ministero
dell'Istruzione, dell'Universit\`a e della Ricerca (MIUR), by
Universit\`a di Torino and by Istituto Nazionale di Fisica Nucleare
within the {\sl Astroparticle Physics Project} (MIUR contract number: PRIN 2008NR3EBK;
INFN grant code: FA51).
NF acknowledges support of the spanish MICINN
Consolider Ingenio 2010 Programme under grant MULTIDARK CSD2009- 00064 (MICINN).}

\appendix

\section{The grid}
\label{app:grid}

\begin{figure*}[tb]
	\centering
	\includegraphics[width=0.69\textwidth,bb = 0 0 720 445]{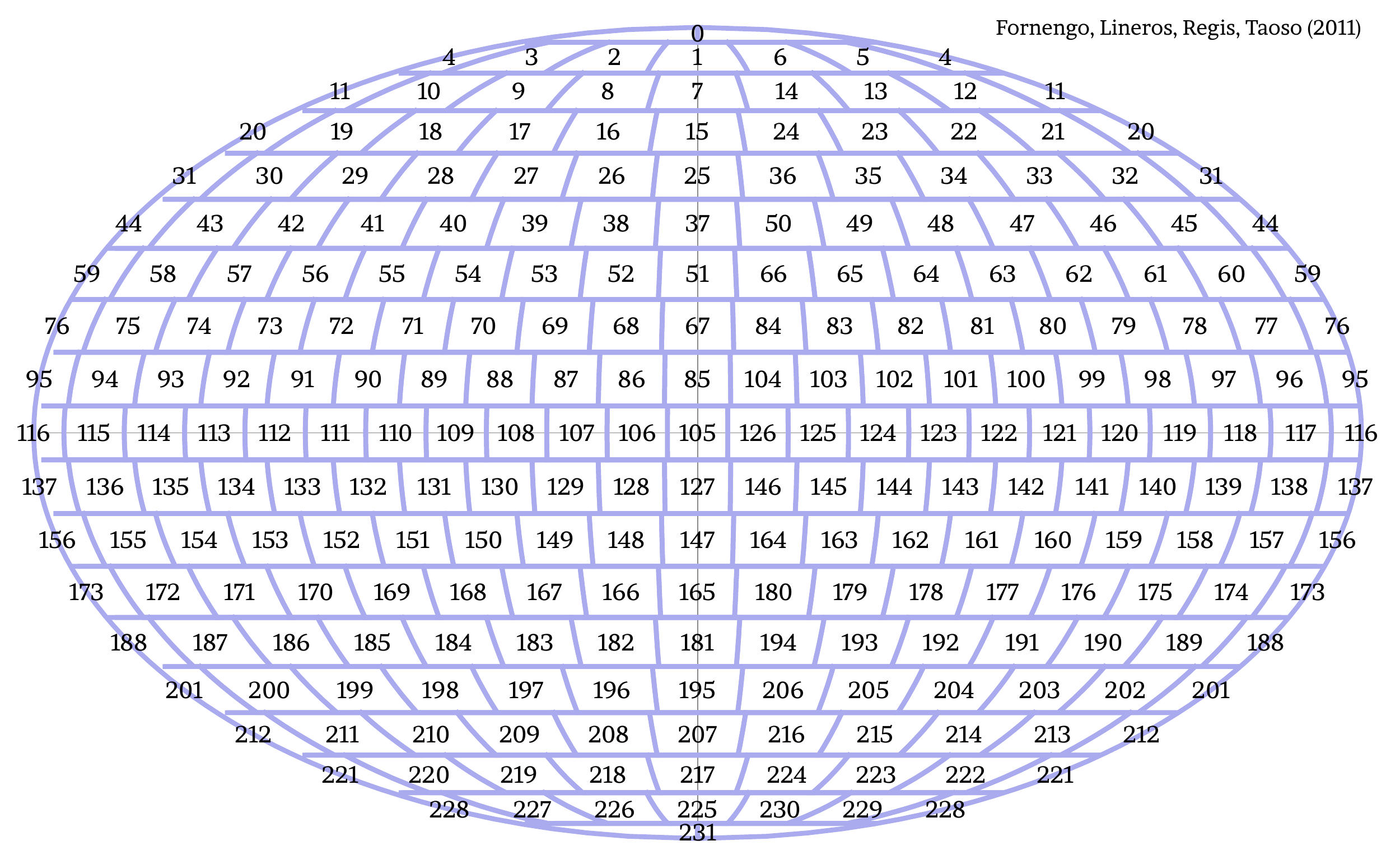}
	\caption{\label{grid} Mollweide projection of the grid used in this work. Patch indexes are also shown. The Galactic center is contained in patch 105.}	
\end{figure*}

The divide the sky with a grid in spherical angles $\theta$ and $\phi$ which depend on the
following 4 parameters: \\
\begin{itemize}
\item $\theta_{\rm{off}}$ sets angular scale of poles.
\item $n_{\theta}$ sets number of $\theta$--belts (without counting polar caps).
\item $n_{\phi}^{\rm{dk}}$ sets number of $\phi$--patches at the galactic plane ($\theta = \pi/2$).
\item $n_{\phi}^{\rm{pl}}$ sets number of $\phi$--patches for the belts closest to the poles.
\end{itemize}

In this work, we choose $\theta_{\rm{off}} = 7.5^{\circ}$, $n_{\theta}= 18$, $n_{\phi}^{\rm{dk}} = 22$, and $n_{\phi}^{\rm{pl}} = 6$. 
This specific configuration produces a grid where the galactic plane ($\theta = \pi/2$) is covered with one belt in $\theta$. 
Moreover, it produces two columns centered on $\phi = 0$ and $\phi = \pi$ covering the galactic center and the galactic anticenter (figure \ref{grid}).\\

The first step is to generate the $\theta$--belts.
Each one is bounded by $\theta_i$ and $\theta_{i+1}$, defined by:
\begin{equation}
	\theta_i = \left\{ 
		\begin{array}{ccl}
		0 & \rm{for} & i =  0  \\
		\displaystyle \theta_{\rm{off}} + \left(\pi - 2 \theta_{\rm{off}}\right)\frac{i-1}{n_{\theta}-1} & \rm{for} & 1 \leq i \leq n_{\theta} \\
		\pi & \rm{for} & i = n_{\theta}+1 
		\end{array}
	\right.
\end{equation}
There are in total $n_{\theta}-1$ belts, that cover the region $\theta \in [\theta_{\rm{off}}, \pi - \theta_{\rm{off}}]$, and 2 patches covering the poles.
Each $\theta$--belt is divided as well into $j^{\rm{max}} + 1$ equally distributed patches in $\phi$:
\begin{equation}
	\phi_{i,j} = \left\{  \begin{array}{ccl} 
		- \phi_{\rm{off}} & \rm{for} & j = 0 \\ & & \\
		\displaystyle \phi_{\rm{off}} + {\left(2 \pi - 2 \phi_{\rm{off}}\right)} \frac{j-1}{j^{\rm{max}}_i - 1} & \rm{for} & 1 \leq j \leq j^{\rm{max}}
	\end{array} \right. \ ,
\end{equation} 
where $\phi_{\rm{off}} = \pi/j^{\rm{max}}_i$  ensures a column of $\phi$--patches centered on $\phi = 0$. 
These parameters also control the value of $j^{\rm{max}}_i$:
\begin{equation}
	j^{\rm{max}}_i = \left\{ \begin{array}{ccl} 
		\displaystyle n_{\phi}^{\rm{pl}} + {\left(n_{\phi}^{\rm{dk}} - n_{\phi}^{\rm{pl}}\right)} \frac{i-1}{\frac{n_{\theta}}{2} - 1} & \rm{for} &  1 \leq i \leq \frac{n_{\theta}}{2} \\ & & \\
		\displaystyle n_{\phi}^{\rm{pl}} + {\left(n_{\phi}^{\rm{dk}} - n_{\phi}^{\rm{pl}}\right)} \frac{n_{\theta} - i}{\frac{n_{\theta}}{2} - 1} & \rm{for} &  \frac{n_{\theta}}{2}  < i \leq n_{\theta}
	\end{array} \right. \ ,
\end{equation}
which linearly interpoles (using division of integer numbers) between $n_{\phi}^{\rm{pl}}$ and $n_{\phi}^{\rm{dk}}$.\\

Then, each skypatch $\Omega_k$ is defined by 4 points in the $\theta$--$\phi$ plane. We follow the convection
\begin{equation}
	\Omega_k = \left(\theta_{\min},\theta_{\max}\right)\times\left(\phi_{\min}, \phi_{\max}\right) \, ,
\end{equation}
where $k$ is just the patch index. 
$k$ is assigned by counting the number of patches going systematically from the North to the South pole by covering first each $\phi$--patch in every $\theta$--belts.\\ 

The Northen polar cap is:
\begin{equation}
	\Omega_0 = \left(0, \theta_{\rm{off}}\right)\times\left(0 , 2\pi \right) \, ,
\end{equation}
and the patches in the North hemisphere are:
\begin{equation}
	\Omega_k = \left(\theta_i, \theta_{i+1}\right)\times\left(\phi_{i,j},\phi_{i,j+1} \right) \, ,
\end{equation}
for $1\leq i \leq n_{\theta}/2$ and $0 \leq j \leq j^{\rm{max}}_i - 1$.\\

The patches in the South hemisphere are 
\begin{equation}
	\Omega_k = \left(\theta_{i-1}, \theta_{i}\right)\times\left(\phi_{i,j},\phi_{i,j+1} \right) \, ,
\end{equation}
for $n_{\theta}/2  + 2 \leq i \leq n_{\theta}/2$ and $0 \leq j \leq j^{\rm{max}}_i - 1$.\\
 
The last patch corresponds to the Southern polar cap:
\begin{equation}
	\Omega_k = \left(\pi - \theta_{\rm{off}}, \pi\right)\times\left(0 , 2\pi \right) \ .
\end{equation}

\section{Synchrotron emission power}
\label{app:synch}
The synchrotron emission power at frecuency $\nu$ produced by a relativistic electron with energy $E$~\cite{Ginzburg:1965} is:
\begin{equation}
\frac{dw}{d\nu}(\nu,B_{\perp}) = \frac{\sqrt{3}\, e^3 B_{\perp}}{m_e c^2} \, F\left(\frac{\nu}{\nu_{c,\perp}}\right)\,,
\end{equation}
where 
\begin{equation}
	\nu_{c,\perp} = \frac{3 e B_{\perp} E^2}{4 \pi m_e^3 c^5} \ {\rm and} \ 	F(x) = x \int_{x}^{\infty} d\zeta \ K_{5/3}(\zeta) \,,
\end{equation}
with $K_{5/3}$ is a modified Bessel function. 
The emission depends only the perperdicular component of the magnetic field ($B_{\perp} = B \sin{\theta}$) with respect to the electron's momentum ($\theta$).\\

In the case of cosmic rays, the dependence on $\theta$ will be softened due to the diffuse (isotropic) propagation of cosmic rays.
In this case, we define the function:
\begin{equation}
	G(x) = \frac{2}{\pi} \int_{0}^{\frac{\pi}{2}} d \theta \sin\theta F\left(\frac{x}{\sin\theta}\right)
\end{equation}
which is the angular average of the $F$ function.
Moreover, the $G$ function can be well fitted by:
\begin{equation}
	G(x) = a x^d \exp\left(-\sqrt{\frac{x}{b}} - \frac{x}{c}\right)\,,
\end{equation}
where the numerical values of parameters $a, b,c,d$ are:
\begin{equation}
	a = 1.60883, b = 1.95886, c = 1.13147, d = 0.33839 \ .
\end{equation}

Using this function, the emission power for an isotropic distribution of electrons is:
\begin{equation}
	\frac{dw}{d\nu} = \frac{\sqrt{3}\, e^3 B}{m_e c^2} \, G\left(\frac{\nu}{\nu_c}\right)\,,
\end{equation}
where $\nu_c = {3 e B E^2}/{(4 \pi m_e^3 c^5)}$ and $B$ is the total magnetic field intensity.\\

\bibliographystyle{JHEP}
\bibliography{./bib/references}

\end{document}